\pdfoutput=1

\documentclass[11pt,twoside,a4paper,cmspaper,final,collab]{cms-tdr}
\def\svnVersion{43a3768-D}\def\svnDate{2020/06/12}\def\cmsCernNoTag{CERN-EP-2019-281}\def\cmsCernDate{\today}\def\cmsMessage{"Published in Machine Learning: Science and Technology as \href{http://dx.doi.org/10.1088/2632-2153/ab9023}{\doi{10.1088/2632-2153/ab9023}.}"}

\begin{document}\cmsNoteHeader{EXO-19-011}

\hyphenation{had-ron-i-za-tion}
\hyphenation{cal-or-i-me-ter}
\hyphenation{de-vices}

\ifthenelse{\boolean{cms@external}}{\providecommand{\cmsLeft}{upper\xspace}}{\providecommand{\cmsLeft}{left\xspace}}
\ifthenelse{\boolean{cms@external}}{\providecommand{\cmsRight}{lower\xspace}}{\providecommand{\cmsRight}{right\xspace}}
\newcommand{\mgluino}{\ensuremath{m_{\PSg}}\xspace}
\newcommand{\lsp}{\PSGczDo}
\newcommand{\mlsp}{\ensuremath{m_{\lsp}}\xspace}
\newcommand{\ctau}{\ensuremath{c\tau_{0}}\xspace}
\newcommand{\mdphi}{\ensuremath{\Delta\phi_\text{min}^\star}\xspace}
\newcommand{\njet}{\ensuremath{N_\text{jet}}\xspace}
\newcommand{\ntag}{\ensuremath{N_\text{tag}}\xspace}
\newcommand{\prob}{\ensuremath{P(\mathrm{LLP}|\ctau)}\xspace}
\newcommand{\pmax}{\ensuremath{P_\text{max}(\mathrm{LLP}|\ctau)}\xspace}
\newcommand{\nsv}{\ensuremath{N_\mathrm{SV}}\xspace}
\newcommand{\edata}{\ensuremath{\epsilon_\text{data}}\xspace}
\newcommand{\emc}{\ensuremath{\epsilon_\mathrm{MC}}\xspace}
\newcommand{\nllp}{\ensuremath{N_\text{LLP}}\xspace}
\newcommand{\ntllp}{\ensuremath{N_\text{LLP}^\text{tag}}\xspace}
\newcommand{\sfac}{\ensuremath{\mathrm{SF}}\xspace}
\newcommand{\mjets}{\ensuremath{\PGm\text{+jets}}\xspace}
\newcommand{\mmjets}{\ensuremath{\PGm\PGm\text{+jets}}\xspace}
\newcommand{\wlnujets}{\ensuremath{\PW({\to}\ell\PGn)\text{+jets}}\xspace}
\newcommand{\zmumujets}{\ensuremath{\PZ/\PGg^*({\to}\PGm\PGm)\text{+jets}}\xspace}
\newcommand{\znunujets}{\ensuremath{\PZ({\to}\PGn\PAGn)\text{+jets}}\xspace}
\newcommand{\tensorflow}{\textsc{TensorFlow}\xspace}
\newcommand{\keras}{\textsc{Keras}\xspace}
\newcommand{\ROOT}{\textsc{root}\xspace}
\newlength\cmsTabSkip\setlength{\cmsTabSkip}{1.0ex}

\title{A deep neural network to search for new long-lived particles
  decaying to jets}

\date{\today}

\abstract{A tagging algorithm to identify jets that are significantly
  displaced from the proton-proton ($\Pp\Pp$) collision region in the
  CMS detector at the LHC is presented. Displaced jets can arise from
  the decays of long-lived particles (LLPs), which are predicted by
  several theoretical extensions of the standard model. The tagger is
  a multiclass classifier based on a deep neural network, which is
  parameterised according to the proper decay length $c\tau_0$ of the
  LLP. A novel scheme is defined to reliably label jets from LLP
  decays for supervised learning. Samples of $\Pp\Pp$ collision data,
  recorded by the CMS detector at a centre-of-mass energy of 13\TeV,
  and simulated events are used to train the neural network. Domain
  adaptation by backward propagation is performed to improve the
  simulation modelling of the jet class probability distributions
  observed in $\Pp\Pp$ collision data. The potential performance of
  the tagger is demonstrated with a search for long-lived gluinos, a
  manifestation of split supersymmetric models. The tagger provides a
  rejection factor of 10\,000 for jets from standard model processes,
  while maintaining an LLP jet tagging efficiency of 30--80\% for
  gluinos with $1\mm \leq c\tau_0 \leq 10\unit{m}$. The expected
  coverage of the parameter space for split supersymmetry is
  presented.}

\hypersetup{
  pdfauthor={CMS Collaboration},
  pdftitle={A deep neural network to search for new long-lived
    particles decaying to jets},
  pdfsubject={CMS},
  pdfkeywords={CMS, split SUSY, long-lived particles, deep neural
    network}
}

\maketitle

\section{Introduction}
\label{sec:intro}

{Machine-learned algorithms are routinely deployed to perform event
  reconstruction, particle identification, event classification, and
  other tasks~\cite{Albertsson:2018maf} when analysing data samples
  recorded by experiments at the CERN LHC. Machine learning techniques
  have been widely adopted to classify a jet, a collimated spray of
  particles that originate from the hadronisation of a parton,
  according to the underlying flavour of the initial
  parton~\cite{Larkoski:2017jix}. For example, jets that originate
  from the hadronisation of {\PQb} quarks (\PQb jets) exhibit
  characteristic experimental signatures that can be exploited by
  dedicated algorithms to identify \PQb jets, in a procedure known as
  {\PQb} tagging. The {\PQb} hadrons, with proper lifetimes of
  $\mathcal{O}(10^{-12}\unit{s})$, typically travel distances of
  approximately 1--10\mm, depending on their momenta, before
  decaying. As a result, charged particle tracks in jets can originate
  from one or more common vertices that may be displaced with respect
  to the proton-proton ($\Pp\Pp$) collision region. Furthermore, the
  impact parameter of each track, defined as the spatial distance
  between the originating $\Pp\Pp$ collision and the track at its
  point of closest approach, can have a significant nonzero value. The
  ATLAS~\cite{Aad:2008zzm} and CMS~\cite{Chatrchyan:2008zzk}
  Collaborations have developed numerous algorithms based on boosted
  decision trees or neural networks to identify \PQb
  jets~\cite{Sirunyan:2017ezt, Aad:2019aic} using the aforementioned
  and other high-level engineered features. The latest {\PQb} tagging
  algorithm developed by the CMS Collaboration is the DeepJet
  tagger~\cite{Stoye_2018, CMS-DP-2018-058}, which is a multiclass
  classifier that discriminates between jets originating from the
  hadronisation of heavy- ({\PQb} or {\PQc}) or light-flavour ({\PQu},
  {\PQd}, and {\PQs}) quarks or gluons ({\Pg}) with unprecedented
  performance. The algorithm is based on a deep neural network (DNN)
  that exploits particle-level information, as well as jet-level
  engineered features used in preceding b tagging
  algorithms~\cite{Sirunyan:2017ezt, Aad:2019aic}.  }

Various theoretical extensions of the standard model
(SM)~\cite{ArkaniHamed:2004fb, Giudice:2004tc, Hewett:2004nw,
  Giudice:1998bp, Barbier:2004ez, Strassler:2006ri, Han:2007ae}
predict the existence of long-lived particles (LLPs) with a proper
lifetime $\tau_{0}$ that can be very different from those of known SM
particles. Consequently, the production and decay of LLPs at the LHC
could give rise to atypical experimental signatures. These
possibilities have led to the development of a broad search programme
at the LHC, based around LLP simplified models~\cite{sms,
  Buchmueller:2017uqu} and novel reconstruction techniques. A
comprehensive review of LLP searches at the LHC can be found in
Ref.~\cite{Alimena:2019zri}.

In this paper, we present the novel application of a DNN to tag (\ie
identify) a jet originating from the decay of an LLP (LLP jet), both
in the presence or absence of a displaced vertex of charged particle
tracks. The DNN is trained and evaluated using a range of signal
hypotheses comprising simplified models of split supersymmetry
(SUSY)~\cite{ArkaniHamed:2004fb, Giudice:2004tc} with $R$
parity~\cite{Farrar:1978xj} conservation. These models assume the
production of gluino ({\PSg}) pairs. The gluino is a long-lived state
that decays to a quark-antiquark ({\PQq}{\PAQq}) pair and a weakly
interacting and massive neutralino (\lsp), which is the lightest SUSY
particle and a dark matter candidate. Simplified models of SUSY that
yield LLPs are widely used as a benchmark for searches in final states
containing jets and an apparent imbalance in transverse momentum,
\ptvecmiss~\cite{Aaboud:2019opc, Aaboud:2019trc, Aaboud:2018aqj,
  Aaboud:2017iio, Sirunyan:2019gut, Sirunyan:2018vlw,
  Sirunyan:2018pwn, Sirunyan:2018vjp, Sirunyan:2017sbs,
  Sirunyan:2017jdo, Khachatryan:2016sfv, Aaij:2017mic}. The search in
Ref.~\cite{Sirunyan:2018vjp} does not explicitly target LLPs, but its
inclusive approach provides sensitivity over a large range of
lifetimes. This search is used later as a performance benchmark for
the methods presented in this paper.

The LLP jet tagger is inspired by the DeepJet approach, albeit with
significant modifications. The DNN extends the multiclass
classification scheme of the DeepJet algorithm to accommodate the LLP
jet class. A procedure to reliably label LLP jets using
generator-level information from the Monte Carlo (MC) programs is
defined. The experimental signature for an LLP jet depends strongly on
the proper decay length \ctau. Hence, a parameterised
approach~\cite{Baldi:2016fzo} is adopted by using \ctau as an external
parameter to the DNN, which permits hypothesis testing using a single
network for models with values of \ctau that span several orders of
magnitude. Furthermore, the jet momenta depend strongly on the masses
\mgluino and \mlsp, and particularly the mass difference $\mgluino -
\mlsp$. The DNN training is performed with simulated event samples
drawn from the full mass parameter space of interest to ensure a broad
generalisation and optimal performance over a range of jet
momenta. The performance of the tagger is also quantified for
simplified models of SUSY with gauge-mediated SUSY breaking
(GMSB)~\cite{Giudice:1998bp} and weak $R$-parity violation
(RPV)~\cite{Barbier:2004ez, Graham:2012th}. Domain adaptation by
backward propagation of errors~\cite{ganin2014unsupervised} is
incorporated into the network architecture to achieve similar
classification performance in simulation and $\Pp\Pp$ collision data,
thus mitigating differences between the two domains of simulation and
data. Differences can arise because of, for example, the limited
precision of the simulation. The $\Pp\Pp$ collision data were recorded
by the CMS detector at a centre-of-mass energy of 13\TeV. The sample,
recorded in 2016, corresponds to an integrated luminosity of
35.9\fbinv.

This paper is organised as follows. Section~\ref{sec:cms} describes
the CMS detector and event reconstruction algorithms.
Sections~\ref{sec:selections} and~\ref{sec:simulation} describe,
respectively, the event samples and simulation software packages used
in this study. Section~\ref{sec:network} describes the LLP jet tagger.
Sections~\ref{sec:validation} and~\ref{sec:performance} demonstrate,
respectively, the validation of the tagger using control samples of
$\Pp\Pp$ collision data and its performance based on simulated
samples. Section~\ref{sec:showcase} presents the expected performance
of the tagger in a search for long-lived gluinos, as well as an \textit{in situ}
determination of a correction to the signal efficiency of the LLP
jet tagger. Section~\ref{sec:summary} provides a summary of this work.

\section{The CMS detector and event reconstruction}
\label{sec:cms}

The central feature of the CMS apparatus is a superconducting solenoid
of 6\unit{m} internal diameter, providing a magnetic field of
3.8\unit{T}. Within the solenoid volume are a silicon pixel and strip
tracker, a lead tungstate crystal electromagnetic calorimeter (ECAL),
and a brass and scintillator hadron calorimeter (HCAL), each composed
of a barrel and two endcap sections.  Forward calorimeters extend the
pseudorapidity ($\eta$) coverage provided by the barrel and endcap
detectors. Muons are detected in gas-ionisation chambers embedded in
the steel flux-return yoke outside the solenoid. At $\eta = 0$, the
outer radial dimension of the barrel section of the tracker, ECAL,
HCAL, and muon subdetector is 1.3, 1.8, 3.0, and 7.4\unit{m},
respectively. A more detailed description of the CMS detector,
together with a definition of the coordinate system used and the
relevant kinematic variables, can be found in
Ref.~\cite{Chatrchyan:2008zzk}.

{The candidate vertex with the largest value of summed physics-object
  $\pt^2$ is taken to be the primary $\Pp\Pp$ interaction vertex,
  where $\pt$ is the transverse momentum. Here, the physics objects
  are the jets, clustered using the jet finding
  algorithm~\cite{Cacciari:2008gp,Cacciari:2011ma} with the tracks
  assigned to candidate vertices as inputs, and the associated
  \ptvecmiss, taken as the negative vector \pt sum of those jets.
}

The particle-flow (PF) algorithm~\cite{CMS-PRF-14-001} aims to
reconstruct and identify each particle (PF candidate) in an event,
with an optimised combination of all subdetector information. In this
process, the identification of the particle type (photon, electron,
muon, charged or neutral hadron) plays an important role in the
determination of the particle direction and energy.
Photons~\cite{Khachatryan:2015iwa} are identified as ECAL energy
clusters not linked to the extrapolation of any charged particle
trajectory to the ECAL. Electrons~\cite{Khachatryan:2015hwa} are
identified as a primary charged particle track and potentially many
ECAL energy clusters corresponding to the extrapolation of this track
to the ECAL and to possible bremsstrahlung photons emitted along the
way through the tracker material. Muons~\cite{Sirunyan:2018fpa} are
identified as tracks in the central tracker consistent with either a
track or several hits in the muon system, and associated with
calorimeter deposits compatible with the muon hypothesis. Charged
hadrons are identified as charged particle tracks neither identified
as electrons, nor as muons. Finally, neutral hadrons are identified as
HCAL energy clusters not linked to any charged hadron trajectory, or
as a combined ECAL and HCAL energy excess with respect to the expected
charged hadron energy deposit. The inclusive vertex finding
algorithm~\cite{Khachatryan:2011wq} is used as the standard secondary
vertex reconstruction algorithm, which uses all reconstructed tracks
in the event with $\pt > 0.8\GeV$ and a longitudinal impact parameter
of greater than 0.3\unit{cm}.

The PF algorithm is able to reconstruct reliably particle candidates
with large displacements.  In the inner tracker volume, charged
particles are identified through the presence of an associated track.
The later iterations of an iterative tracking
procedure~\cite{Chatrchyan:2014fea} explicitly target the most
displaced tracks.  At larger displacements, beyond the tracking
volume, the energy and direction of displaced particles are solely
determined from measurements in the calorimeter systems.  In the case
that an LLP decay occurs in the muon systems, a collimated spray of
muon-track stubs is reconstructed. These stubs are ignored by the PF
algorithm, as they cannot be reliably distinguished from hadronic
showers that are not completely contained in the calorimeters
(so-called hadronic punch-through).

Jets are clustered from PF candidates using the anti-\kt
algorithm~\cite{Cacciari:2008gp, Cacciari:2011ma} with a distance
parameter of 0.4. Additional $\Pp\Pp$ interactions within the same or
nearby bunch crossings (pileup) can contribute additional tracks and
calorimetric energy depositions to the jet momentum. To mitigate this
effect, charged particles identified to be originating from pileup
vertices are discarded~\cite{Cacciari:2007fd} and an offset
correction~\cite{Khachatryan:2016kdb} is applied to correct for
remaining contributions. In this study, jets are required to satisfy
$\pt > 30\GeV$ and $\abs{\eta} < 2.4$, and are subject to a set of
loose identification criteria~\cite{CMS:2017wyc} to reject anomalous
activity from instrumental sources, such as detector noise. These
criteria ensure that each jet contains at least two PF candidates and
at least one charged particle track, the energy fraction attributed to
charged- and neutral-hadron PF candidates is nonzero, and the fraction
of energy deposited in the ECAL attributed to charged and neutral PF
candidates is less than unity.

The most accurate estimator of \ptvecmiss is computed as the negative
 vector \pt sum of all the PF candidates in an event, and its magnitude
is denoted as \ptmiss~\cite{Sirunyan:2019kia}.  The \ptvecmiss is
modified to account for corrections to the energy
scale~\cite{Khachatryan:2016kdb} of the reconstructed jets in the
event. Anomalous high-\ptmiss events can be due to a variety of
reconstruction failures, detector malfunctions, or noncollision
backgrounds.  Such events are rejected by event filters that are
designed to identify more than 85--90\% of the spurious high-\ptmiss
events with a mistagging rate of less than 0.1\% for
genuine events~\cite{Sirunyan:2019kia}.

Events of interest are selected using a two-tiered trigger
system~\cite{Khachatryan:2016bia}. The first level, composed of custom
hardware processors, uses information from the calorimeters and muon
detectors alone, whereas a version of the full event reconstruction
software optimised for fast processing is performed at the second
level, which runs on a farm of processors.

\section{Event selection and sample composition}
\label{sec:selections}

Split SUSY, as characterised by the simplified models discussed in
this paper, would reveal itself in events containing jets, significant
\ptmiss from undetected neutralinos, and an absence of photons and
leptons. Candidate signal events are required to satisfy a set of
selection requirements that define a signal region (SR). Conversely,
event samples that are enriched in the same background processes that
populate the SR, while being depleted in contributions from SUSY
processes, are identified as control regions (CRs).

In this analysis, CRs are used to assess the residual differences in
tagger performance between data and simulation. The CRs are chosen to
have an SM background composition similar to that of the SR. This
permits a validation that the simulated event samples can accurately
predict the SM background yields without significant systematic bias.

Candidate signal events in the SR are required to satisfy the
following set of selection requirements. Events are required to
contain at least three jets, as defined in
Section~\ref{sec:cms}. Events are vetoed if they contain at least one
electron (muon), isolated from other activity in the event, that
satisfies $\pt > 15\,(10)\GeV$, $\abs{\eta} < 2.4$, and loose
identification criteria~\cite{Khachatryan:2015hwa,
  Chatrchyan:2012xi}. The mass scale of each event is estimated from
the scalar \pt sum of the jets, $\HT = \sum_{i}^{\text{jets}}\pt^{i}$,
which is required to be larger than $300\GeV$. An estimator for
\ptmiss is given by the magnitude of the vector \pt sum of the
jets, $\mht = \abs{\sum_{i}^{\text{jets}}\vec{p}^{i}_\mathrm{T}}$,
which is required to be larger than $300\GeV$. Events in the SR can be
efficiently recorded with a trigger condition that requires the
presence of a single jet with $\pt > 80\GeV$, $\mht > 120\GeV$, and
$\ptmiss > 120\GeV$.

Following these selections, the dominant contribution to the SM
background comprises multijet events produced via the strong
interaction, a manifestation of quantum chromodynamics (QCD). The
multijet contribution is reduced to a negligible level using the
following three criteria. Events containing at least one jet that
satisfies $\pt > 50\GeV$ and $2.4 < \abs{\eta} < 5$ are vetoed to
ensure an adequate resolution performance for the \mht
variable. Events are required to satisfy $\mht/\ptmiss < 1.25$, which
mitigates the rare circumstance in which several jets with \pt below
the aforementioned 30\GeV threshold and collinear in $\phi$ lead to
large values of \mht relative to \ptmiss. The minimum azimuthal
separation between each jet and the vector \pt sum of all other jets
in the event, denoted \mdphi~\cite{2011196}, is required to be greater
than 0.2 radians.

The remaining background events in the SR are dominated by
contributions from processes that involve the production of high-\pt
neutrinos in the final state, such as the associated production of
jets and a {\PZ} boson that decays to {\Pgn}{\cPagn}. A further
significant background contribution arises from events that contain a
{\PW} boson that undergoes a leptonic decay, \wlnujets, where the
charged lepton ($\ell \equiv \Pe$, $\PGm$ or $\PGt$) is outside the
experimental acceptance, or is not identified, or is not
isolated. Hence, a substantial contribution is also expected from the
production of single top quarks and top quark-antiquark pairs
(\ttbar), both of which can lead to a final state containing one
leptonically decaying {\PW} boson and at least one {\PQb}
jet. Residual contributions from rare SM processes, such as diboson
production or the associated production of \ttbar and a vector or
scalar boson, are not considered in this study.

Two CRs are used to assess differences in the performance of the
tagger when using simulated events or $\Pp\Pp$ collision data. The CRs
are defined in terms of leptons and jets that satisfy $\abs{\eta} < 2.4$ and
the \pt requirements defined below. The single muon (\mjets) CR is
required to contain exactly one muon satisfying $\pt > 26\GeV$. The
dimuon (\mmjets) CR is required to contain a second muon that
satisfies $\pt > 15\GeV$. The muons are required to be isolated from
other activity in the event and satisfy identification
criteria~\cite{Chatrchyan:2012xi}. Events containing additional
electrons (muons) with $\pt > 15 (10)\GeV$ and satisfying looser
identification criteria, are vetoed. Both CRs must contain at least
two jets satisfying $\pt > 30\GeV$. Events in the \mjets and \mmjets
CRs are required, respectively, to satisfy $\ptmiss > 150\GeV$ and
$\pt(\PGm\PGm) > 100\GeV$, where the latter variable is the magnitude
of the vector \pt sum of the two muons. The \mjets CR comprises, in
approximately equal measure, events from the associated production of
jets and a {\PW} boson, and single top quark and \ttbar
production. The \mmjets CR contains Drell--Yan $(\PQq\PAQq \to
\PZ/\gamma^* \to \PGmpm\PGmmp)$ events with subdominant contributions
from \ttbar and $\PW\PQt$-channel single top quark production.  Events
in both CRs are efficiently recorded with a trigger condition that
requires the presence of a single isolated muon that satisfies $\pt >
24\GeV$.

\section{Monte Carlo simulation}
\label{sec:simulation}

The DNN is trained to predict the jet class using supervised learning,
which relies on generator-level information from MC programs. Various
simulated event samples are also used during the evaluation of the DNN
to benchmark the performance of the tagger.

Split SUSY predicts the unification of the gauge couplings at high
energy~\cite{Dimopoulos:1981yj, Ibanez:1981yh, Marciano:1981un} and a
candidate dark matter particle, the neutralino. Apart from a low mass
scalar Higgs boson, assumed in this model to be the state observed at
125\GeV~\cite{Aad:2012tfa, Chatrchyan:2012ufa}, only the fermionic
SUSY particles may be kinematically accessible at the LHC. All other
SUSY particles are assumed to be ultraheavy. The gluino is only able
to decay through the highly virtual squark states. Hence, the gluino
hadronises and forms a bound state with SM particles called an $R$
hadron. The $R$ hadron can travel a significant distance before the
gluino undergoes a three-body decay to $\PQq\PAQq\lsp$, depending on
\ctau.

The split SUSY simplified models are defined by three parameters:
\ctau and the masses of the gluino \mgluino and the neutralino
\mlsp. The following model parameter space is considered by the
search: $600 < \mgluino < 2400\GeV$, $\mgluino - \mlsp > 100\GeV$, and
$10\mum < \ctau < 10\unit{m}$. The lower and upper bounds of the \ctau
range are motivated by the $\mathcal{O}(10\mum)$ position resolution
of the tracker subdetector~\cite{Chatrchyan:2014fea} and the physical
dimensions of the CMS detector, respectively. The tagger performance
is also assessed using two split SUSY benchmark models that feature:
an ``uncompressed'' mass spectrum, with a large mass difference
between the gluino and \lsp, $(\mgluino,\mlsp) = (2000,0)\GeV$; or a
``compressed'' spectrum that is nearly degenerate in mass,
$(1600,1400)\GeV$. The value of \ctau for both models is defined in
the text on a case-by-case basis.

Two benchmark models of GMSB and RPV SUSY are also considered to
demonstrate the generalisation of the DNN. A GMSB-inspired model
assumes a long-lived gluino, with a mass of $2500\GeV$ and $\ctau =
1\mm$ or 1\unit{m}, that decays to a gluon and a light gravitino of
mass 1\unit{keV}. Again, all other SUSY particles are assumed to be
ultraheavy and decoupled from the interaction. An RPV-inspired model
assumes the production of top squark-antisquark pairs. The decay of
the long-lived top squark, with a mass of 1200\GeV and $\ctau = 1\mm$
or 1\unit{m}, to a bottom quark and a charged lepton is suppressed
through a small $R$ parity violating coupling.

Samples of simulated events are produced with several MC generator
programs. For the split SUSY simplified models, the associated
production of gluino pairs and up to two additional partons are
generated at leading-order (LO) precision in QCD with the \MGvATNLO
2.2.2~\cite{Alwall:2014hca} program. The decay of the gluino is
performed with the \PYTHIA 8.205~\cite{Sjostrand:2014zea} program. The
\textsc{rhadrons} package within the \PYTHIA program, steered
according to the default parameter settings, is used to describe the
formation of $R$ hadrons through the hadronisation of
gluinos~\cite{Fairbairn:2006gg, Kraan:2004tz, Mackeprang:2006gx}. A
similar treatment is performed for the GMSB- and RPV-inspired
benchmark models.

The \MGvATNLO event generator is used to produce samples of \wlnujets,
\znunujets, and \zmumujets events at next-to-leading-order (NLO)
precision in QCD. The samples of \wlnujets events are generated with
up to two additional partons at the matrix element level and are
merged with jets from the subsequent \PYTHIA parton shower simulation
using the FxFx scheme~\cite{Frederix:2012ps}. The \POWHEG
v2~\cite{Nason:2004rx, Frixione:2007vw, Alioli:2010xd} event generator
is used to simulate \ttbar production~\cite{Frixione:2007nw} and the
$t$-channel~\cite{Alioli:2009je} and
$\PW\PQt$-channel~\cite{Re:2010bp} production of single top quarks at
NLO accuracy. Multijet events are simulated at LO accuracy using
\PYTHIA.

The theoretical production cross sections for SM processes are
calculated with NLO and next-to-NLO precision~\cite{wphys, fewz,
  top++, Alioli:2009je, Re:2010bp, Alwall:2014hca}. The production
cross sections for gluino and top squark-antisquark pairs are
calculated with NLO plus next-to-leading-logarithmic
precision~\cite{Borschensky:2014cia}.

The \PYTHIA program with the CUETP8M1 tune~\cite{Khachatryan:2015pea}
is used to describe parton showering and hadronisation for all
simulated samples except top quark-antiquark production, which used
the CUETP8M2T4 tune~\cite{CMS-PAS-TOP-16-021}. The NNPDF3.0 LO and NLO
parton distribution functions~\cite{Ball:2014uwa} are used with the LO
and NLO event generators, respectively. Minimum bias events are
overlaid with the hard scattering event to simulate pileup
interactions, with the multiplicity distribution matched to that
observed in $\Pp\Pp$ collision data. The resulting events undergo a
full detector simulation using the
\GEANTfour~\cite{Agostinelli:2002hh} package.  The analysis described
in Section~\ref{sec:showcase} focuses on identifying the decay
products of the LLP, characterised by tracks with large impact
parameters, calorimeter deposits, and secondary vertices. Interactions
of $R$ hadrons with the detector material that can result in signatures
with short tracks or anomalous energy loss are therefore not
considered in this study.

\section{The LLP jet algorithm}
\label{sec:network}

In this section, the DNN architecture and its technical implementation
are presented. The use of a single parameterised DNN for hypothesis
testing and the application of domain adaptation (DA) to samples of
$\Pp\Pp$ collision data recorded by the CMS experiment are presented
for the first time.

\subsection{Jet labelling}
\label{sec:labels}

{Generator-level information from MC programs is often used to label a
  jet according to its initiating parton for supervised learning. A
  standard procedure known as ``ghost''
  labelling~\cite{Cacciari:2007fd} determines the jet flavour by
  clustering the reconstructed final-state particles and the
  generator-level {\PQb} and {\PQc} hadrons into jets.  Only the
  directional information of the four-momentum of the generator-level
  (ghost) hadron is retained to prevent any modification to the
  four-momentum of the corresponding reconstructed jet. Jets
  containing at least one ghost hadron are assigned the corresponding
  flavour label, with {\PQb} hadrons preferentially selected over
  {\PQc} hadrons. Similarly, labels are defined for jets originating
  from gluons ({\Pg}) or light-flavour ({\PQu}{\PQd}{\PQs}) quarks.  }

\begin{figure}[!th]
  \centering
  \includegraphics[width=0.48\textwidth]{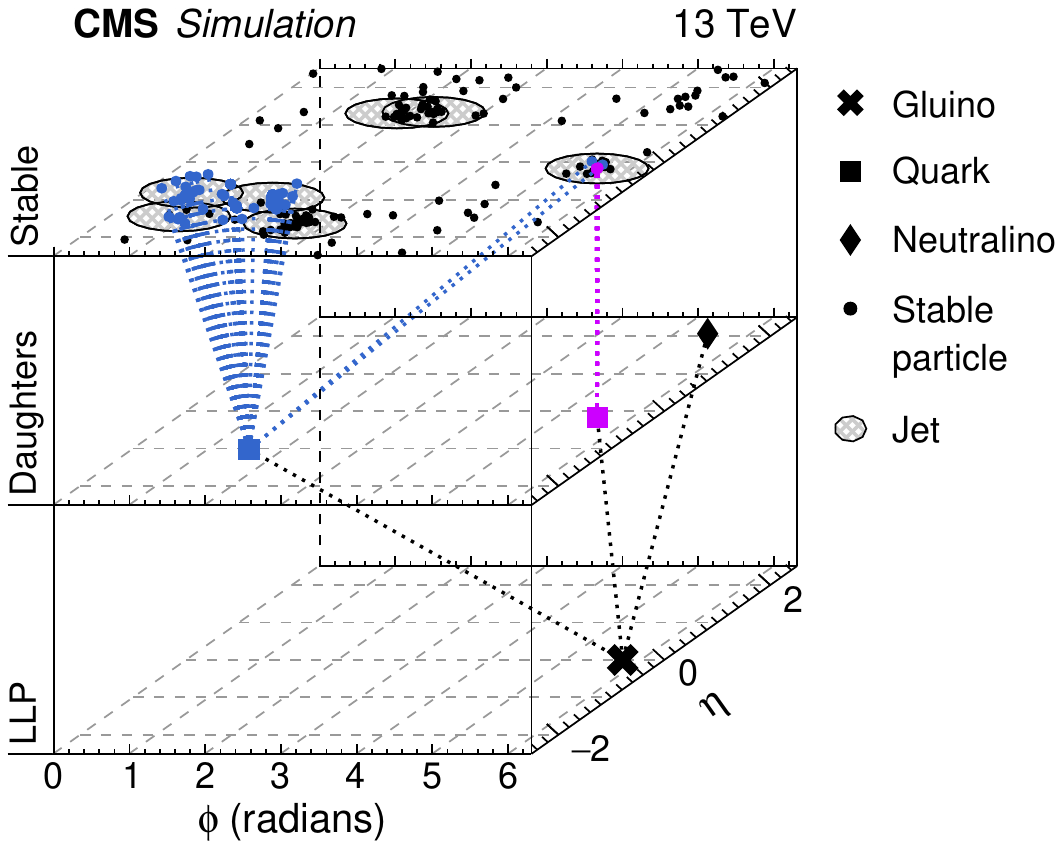}\hspace{0.03\textwidth}
  \includegraphics[width=0.48\textwidth]{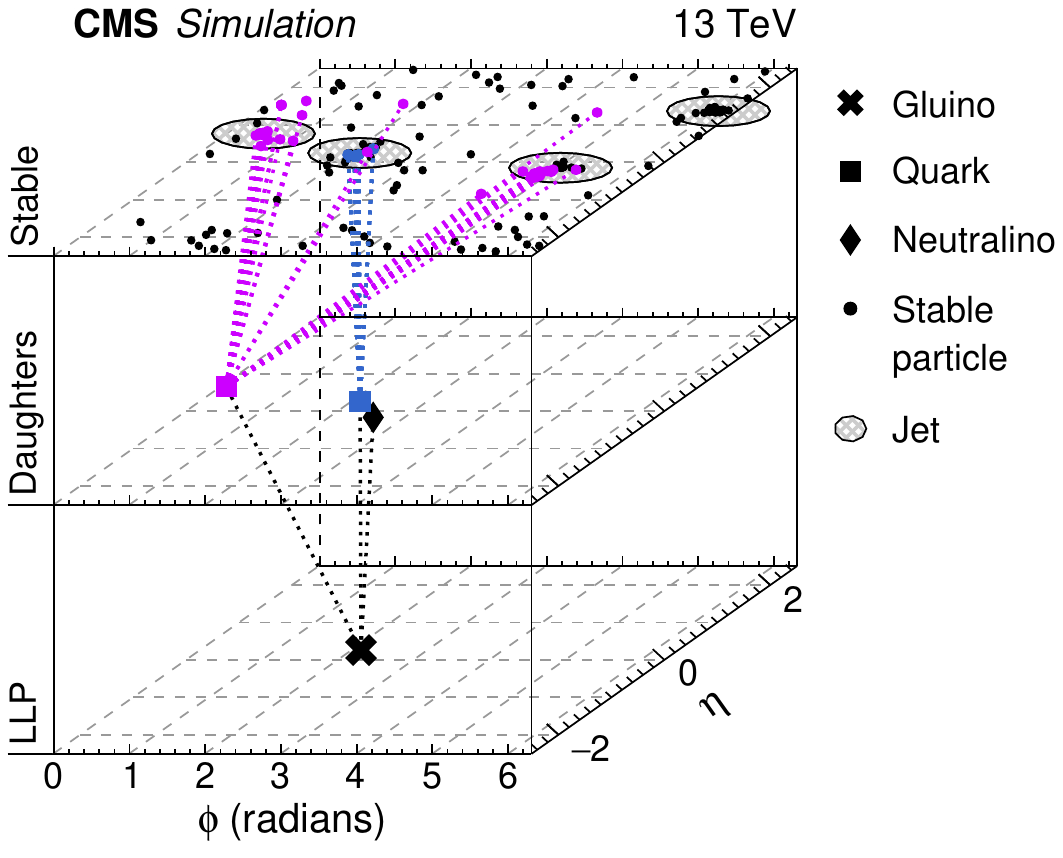}
  \caption{Two example $\PSg\to\PQq\PAQq\lsp$ decay chains,
    constructed from information provided by the
    \MGvATNLO~\cite{Alwall:2014hca} and
    \PYTHIA~\cite{Sjostrand:2014zea} programs. The positions of
    various particles in the $\eta$--$\phi$ plane are shown: the LLP
    ({\PSg}) and its daughter particles ($\PQq\PAQq\lsp$) are shown in
    the lower and middle planes, respectively; the upper plane depicts
    the location of the stable particles after hadronisation, with
    shaded ellipses overlaid to indicate the reconstructed jets. Each
    quark and its decay is assigned a unique colour. The dotted lines
    indicate the links between parent and daughter
    particles. \label{fig:tagger-displaced-label}}
\end{figure}

The LLP jet tagger adopts the ghost labelling approach for jets
originating from SM background processes. However, a complication
arises when applying ghost labelling to the jets originating from
{\PSg} decays. The quark and antiquark produced in the gluino decay
can interact with each other, potentially leading to one or more jets
that do not point in the same direction as the quarks. Two examples of
$\PSg\to\PQq\PAQq\lsp$ decays are shown in
Fig.~\ref{fig:tagger-displaced-label}. For each example, the
final-state particles resulting from the hadronisation of one of the
quarks are sufficiently diffuse that they are clustered into multiple
distinct jets. By definition, ghost tagging cannot account for
multiple jets originating from a single ghost particle, and it may
even fail to associate the ghost particle with any of the jets if the
jets are sufficiently distanced in $\eta$--$\phi$ space.

An alternative labelling scheme is defined for jets originating from
gluino decays, which can be extended to other LLP decays. All stable
SM particles are grouped according to their simulated vertex position
and linked to one of the quark daughters from the LLP decay. All
stable SM particles, except neutrinos, are clustered into
generator-level jets using the anti-\kt
algorithm~\cite{Cacciari:2008gp, Cacciari:2011ma} with a distance
parameter of 0.4. Given that constituent particles in a jet may
originate from different vertices, the momentum $\vec{p}$ of a given
jet is shared between vertices according to the vectorial momentum sum
$\sum_{i}\vec{p}_{i}$ of the constituent particles $i$ in a jet that
share the same generator-level vertex $v$. Per jet, the
jet-vertex shared momentum fraction $f_{v}$ with respect to vertex $v$
is then defined as
\begin{linenomath}
  \begin{equation}
    f_{v}=\frac{\left( \sum_{i} \vec{p}_{i} \middle|\, i \in v\right)
      \cdot \vec{p}}{\vec{p}^2}, \quad f_{v}\in[0;\,1], \quad
    \sum_{v}^\mathrm{vertices} f_{v} = 1.
  \end{equation}
\end{linenomath} {Each jet is then associated to the vertex $v$ from
  which the majority of its momentum originates, \ie $\hat{v} =
  \text{argmax}(f_{v})$. This criterion prevents the coincidental
  association of jets containing very few or very soft particles from
  a gluino decay to a vertex for which the majority of the constituent
  particles stem from initial-state radiation or the underlying event.
  A reconstructed jet is uniquely associated with a generator-level
  jet and they adopt the same label if their axes are aligned within a
  cone $\Delta R = \sqrt{\smash[b]{(\Delta\phi)^2 + (\Delta\eta)^2}} =
  0.4$. The LLP label is prioritised over the other jet labels to
  prevent ambiguities. Jets from the split SUSY samples that are not
  labelled as LLP by this scheme can still comprise a nonnegligible
  fraction of displaced particles and are thus discarded to prevent
  class contamination. Non-LLP jets for the DNN training are instead
  taken from simulated samples of SM backgrounds. Applying an
  artificial 20\% contamination of the LLP jet class from pileup jets,
  to test the robustness of the labelling procedure, leads to no
  discernible effect on the tagger performance given the statistical
  precision of the study.  }

\subsection{Deep neural network architecture to predict the jet class}

\begin{figure*}[!th]
  \centering
  \includegraphics[width=0.83\textwidth]{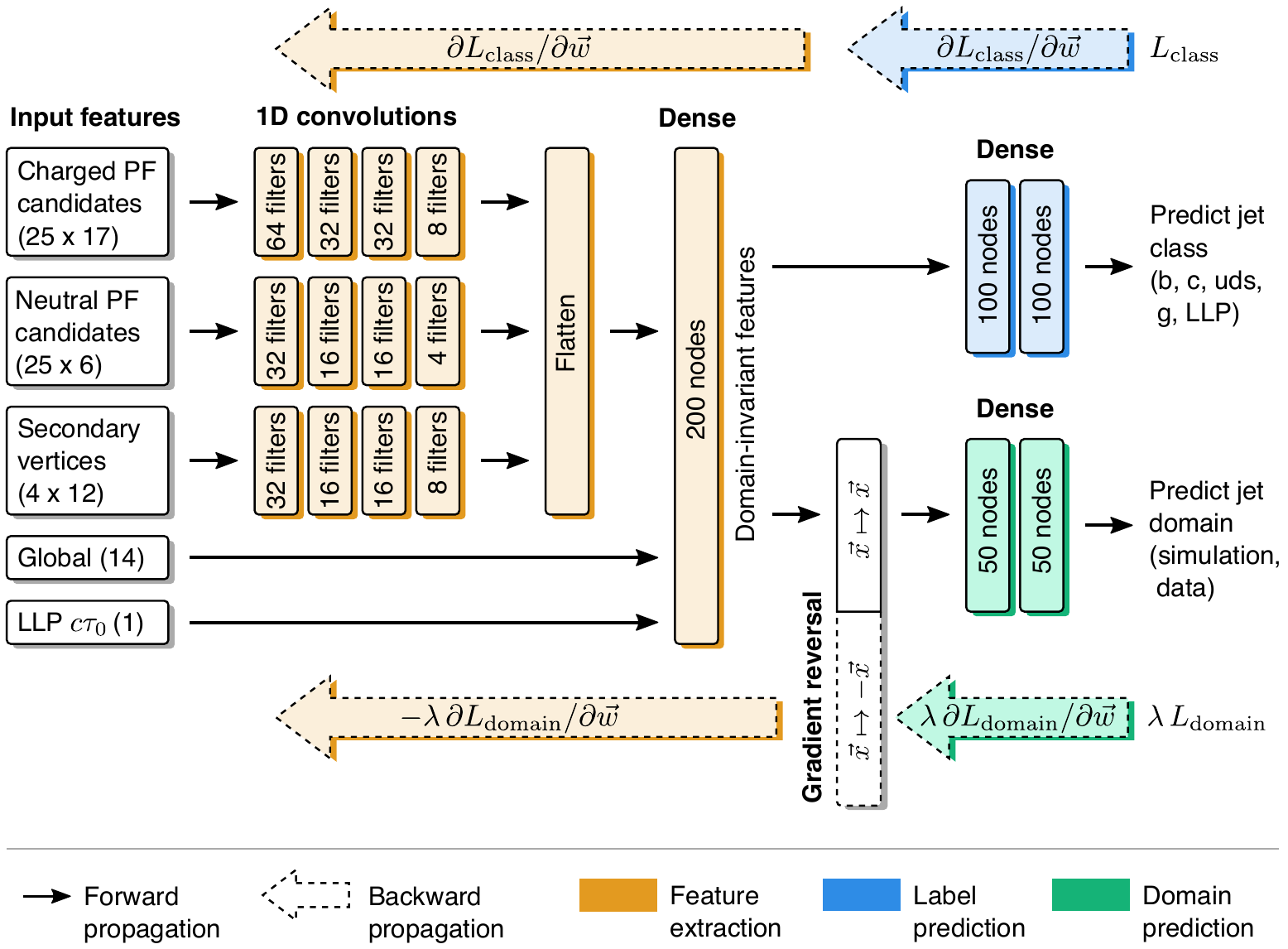}
  \caption{An overview of the DNN architecture, which comprises
    convolutional and dense layers; the numbers of filters and nodes
    are indicated. Dropout layers and activation functions are not
    shown. The input features are grouped by object type and ($m
    \times n$) indicates the maximum number of objects ($m$) and the
    number of features per object ($n$). The gradients of the class
    $(L_\text{class})$ and domain $(L_\text{domain})$ losses with
    respect to the weights $\vec{w}$, used during backward
    propagation, are shown.\label{fig:tagger_network}}
\end{figure*}

The architecture used to predict the jet class, inspired by the
DeepJet algorithm, is shown in Fig.~\ref{fig:tagger_network}.  The DNN
considers approximately 600 input features, which can be grouped into
four categories: up to 25 charged\,(neutral) PF candidates, each
described by 17\,(6) features and ordered by impact parameter
significance (transverse momentum); up to four secondary vertices
(ordered by transverse displacement significance), each described by
12 features; and 14 global features associated with the jet. The
network also considers jets containing zero reconstructed secondary
vertices. Zero padding is used to accommodate the variable numbers of
PF candidates and secondary vertices.

Each charged PF candidate is described by the following features: the
\pt relative and perpendicular to the jet axis, the $\Delta\eta$ with
respect to the jet axis, the track quality, and the transverse and
three-dimensional impact parameters (and their significances) of the
track. Each neutral PF candidate is described by its energy, the
fractions of its energy deposited within the ECAL and HCAL
subdetectors, the compatibility with the photon hypothesis, the
compatibility with the pileup hypothesis as determined by the PUPPI
algorithm~\cite{Bertolini:2014bba, CMS:2019kuk}.  Charged and neutral
PF candidates are also described by the collinearity with respect to
the jet axis and the nearest secondary vertex. The features that
describe each reconstructed secondary vertex include the
three-dimensional displacement (and significance) with respect to the
primary pp collision vertex, the number of associated tracks, and the
following quantities determined from the four-momenta of the
associated tracks: \pt, the $\Delta\eta$ with respect to the jet axis,
and the invariant mass. The global jet features comprise the jet
momentum and pseudorapidity, the number of constituent PF candidates,
the number of reconstructed secondary vertices, and several high-level
engineered features used by the CSV b tagging
algorithm~\cite{Sirunyan:2017ezt}.

Four sequential layers of one-dimensional convolutions with a kernel
size of one are used, with each layer comprising 64, 32, 16, 8, or 4
filters depending on the group of input features. Per particle
candidate or vertex, each convolutional layer transforms the features
from the preceding layer according to its filter size. By choosing a
small filter size for the final layer, the overall operation can be
viewed as a compression. After each layer, a leaky rectified linear
(LeakyReLU)~\cite{glorot2011deep} activation function is
used. Dropout~\cite{srivastava2014dropout} layers are interleaved
throughout the network with a $10\%$ dropout rate to mitigate
overfitting. After the final convolutional layer, the compressed
feature vectors are flattened and concatenated along with the global
jet features. The input parameter \ctau shown in
Fig.~\ref{fig:tagger_network} is described in Section~\ref{sec:ctau}.

The resulting feature vector is fed into a multilayer perceptron, a
series of dense layers comprising 200, 100, or 50 nodes. The softmax
activation function is used in the last layer. Categorical cross
entropy is used for the loss function to predict the jet class
probability. The DNN provides an estimate of the probabilities for the
following jet classes: LLP jet, {\PQb} or {\PQc} jet,
{\PQu}{\PQd}{\PQs} (light-flavour quarks) jet, and {\Pg} jet. The
latter two classes are frequently combined when evaluating the network
performance to give a light-flavour ({\PQu}{\PQd}{\PQs}{\Pg}) jet
class.

\subsection{Network parameterisation according to
  \texorpdfstring{\ctau}{the proper decay length}}
\label{sec:ctau}

The experimental signature for a jet produced in a gluino decay
depends strongly on \ctau. Information from all CMS detector systems
is available if the gluino decay occurs promptly, in the vicinity of
the $\Pp\Pp$ collision region, while information can be limited if the
decay occurs in the outermost detector systems. Hence, \ctau is
introduced as an input parameter to the dense network, as indicated in
Fig.~\ref{fig:tagger_network}. This parameterised
approach~\cite{Baldi:2016fzo} allows for hypothesis testing with a
single network for values of \ctau that span six orders of magnitude:
$10\mum < \ctau < 10\unit{m}$. During the network evaluation for jets
from both signal and SM processes, the \ctau value is given by the
signal hypothesis under test.

\subsection{Domain adaptation by backward propagation}

The simulated event samples are of limited accuracy and do not
exhaustively reproduce all features observed in the pp collision
data. Hence, a neural network may produce different identification
efficiencies when evaluated on simulated samples and $\Pp\Pp$
collision data if the training of the neural network relies solely on
simulation. Domain adaptation is a technique that attempts to
construct a feature representation that is invariant with respect to
the domain from which the features are obtained. In this study, the
domain is either $\Pp\Pp$ collision data or simulation, and the
domain-invariant representation is obtained from the output of the
feature extraction subnetwork, as indicated in
Fig.~\ref{fig:tagger_network}. Hence, for the subsequent task of
classifying jets, a similar performance is expected for both simulated
events and $\Pp\Pp$ collision data~\cite{Ryzhikov_2018}.

In this paper, we use DA by backpropagation of
errors~\cite{ganin2014unsupervised}. To achieve this, an additional
branch is added after the first dense layer of 200 nodes. Binary cross
entropy is used for the loss function to predict the jet domain
probability.  During training, the combined loss $L_\text{class} +
{\lambda}L_\text{domain}$ is minimised, where $\lambda$ is a
hyperparameter that controls the magnitude of the jet domain loss. A
gradient reversal layer is inserted in the domain branch directly
after the first dense layer. This special layer is only active during
backward propagation and reverses the gradients of the domain loss
$L_\text{domain}$ with respect to the network weights $\vec{w}$ in the
preceding layers. This forces the feature extraction subnetwork to
focus less on domain-dependent features because its weights
effectively minimise $L_\text{class} - {\lambda}L_\text{domain}$
instead. Only features that are common to both pp collision data and
simulation are retained. For a full mathematical description of the
gradient reversal layer, see Ref.~\cite{ganin2014unsupervised}.

\subsection{Deep neural network training}

Supervised training of the DNN is performed to predict the jet class
and domain. The Adam optimiser~\cite{adam} is used to minimise the
loss function with respect to the network parameters. The DNN training
relies on simulated events of gluino production from split SUSY
models, and multijet and \ttbar production. Jets in these samples that
have an uncorrected \pt greater than 20\GeV and satisfy the loose
requirements outlined in Section~\ref{sec:cms} are considered for the
DNN training. Approximately 20~million jets are used. Event samples of
split SUSY models are utilised over a wide range of (\mgluino, \mlsp)
hypotheses to ensure adequate generalisation of the DNN over the model
parameter space of interest. The models considered by the DNN training
have seven \ctau values that differ by factors of ten and span the
range $10\mum < \ctau < 10\unit{m}$. For the jet domain prediction,
1.2 million jets are drawn from the pp collision data, as well as from
simulated \wlnujets and \ttbar events that are weighted according to
their respective SM cross sections; all events are required to satisfy
the \mjets CR requirements.

The DNN training comprises a few tens of cycles over the full set of
event samples (epochs). Each epoch is batched into subsamples of
10\,000 jets containing approximately 2000 jets from each class: LLP,
{\PQb}, {\PQc}, {\PQu}{\PQd}{\PQs}, and {\Pg}. For each batch, the LLP
jets are sampled randomly from split SUSY events that are generated
according to the full (\mgluino, \mlsp, \ctau) parameter space of
interest. The jets from SM processes are drawn randomly (from a larger
sample) to obtain the same binned \pt and $\eta$ distributions
available for the LLP jet class. This resampling technique is done to
ensure an adequate generalisation that is largely independent of
kinematical features related to the physics process. The \ctau values
assigned to the SM jets are generated randomly according to the \ctau
distribution obtained per batch for the LLP jet class.

The training of the domain branch uses batches of 10\,000 jets, drawn
from samples of \mjets data and simulated \wlnujets and \ttbar
events. The DNN is trained using the class and domain batches
simultaneously. Events in the domain batch are assigned the same \ctau
values as used by the class batch. For the domain batch, only the six
highest \pt jets are used, and jets may be reused multiple times per
epoch.

The DNN is initially trained to predict only the jet class to
determine the optimal scheduling of the learning rate $\alpha$, which
decays from an initial value of 0.01 according to $\alpha = 0.01 / ( 1
+ \kappa n )$ where $n$ is the epoch number and $\kappa$ is the decay
constant. The classifier performance is optimal for $\kappa = 0.1$ and
only weakly dependent on $\kappa$. The DNN is then trained to predict
both the jet class and domain, and the $\lambda$ hyperparameter is
increased according to $\lambda = \lambda_0[ 2 / (1 + \re^{-0.2n}) - 1
]$ with $\lambda_0 = 30$, such that $\lambda$ increases from 0 to
$0.9\lambda_0$ after 15 epochs. The parameterisations used to evolve
the values of the $\alpha$ and $\lambda$ hyperparameters during the
DNN training were chosen from several trials to ensure reproducible
and optimal performance.

\subsection{Workflow}

\begin{figure*}[!th]
  \centering
  \includegraphics[width=0.83\textwidth]{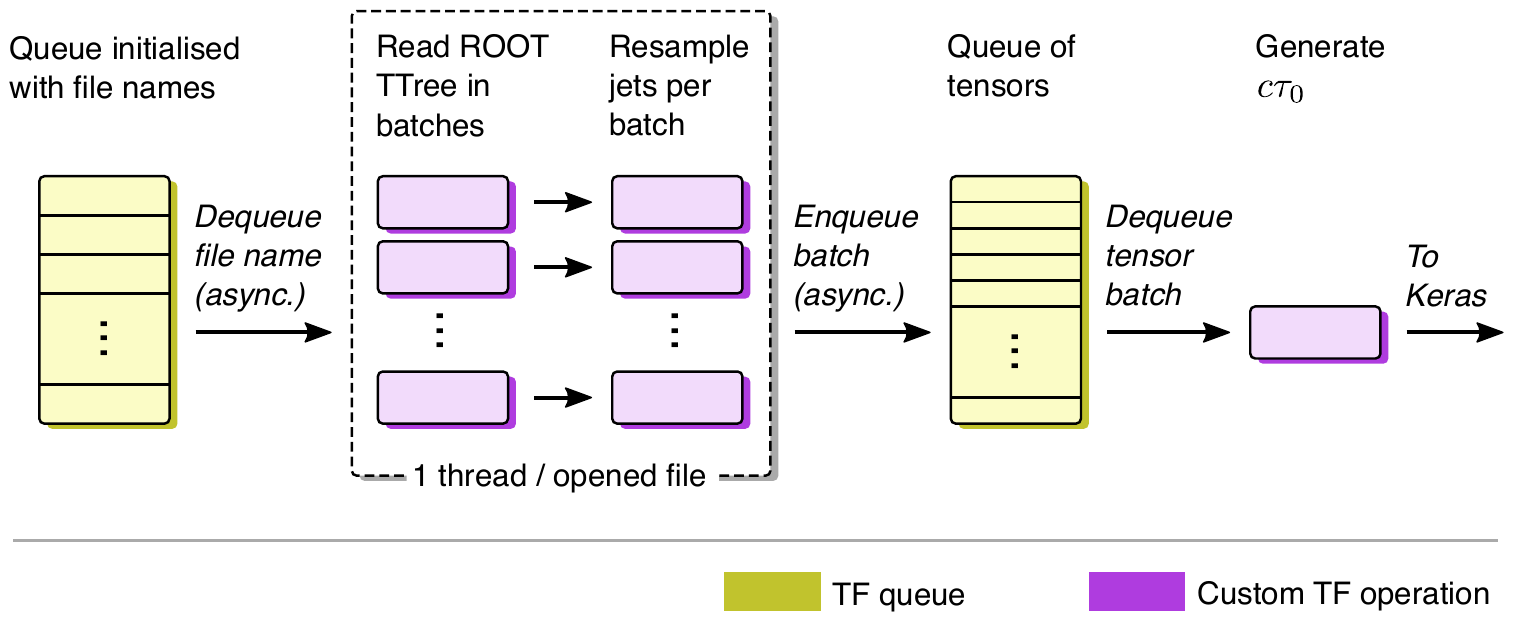}
  \caption{A schematic of the input pipeline for training the DNN,
    which uses the \tensorflow (TF) queue system with custom operation
    kernels for reading \ROOT trees from disk, (\pt, $\eta$)
    resampling for SM jets, and generating random \ctau values for
    jets from SM backgrounds and data. \label{fig:tagger_queues}}
\end{figure*}

{The \keras~v2.1.5~\cite{chollet2015} software package is used to
  implement the DNN architecture. The \tensorflow
  v1.6~\cite{tensorflow2015-whitepaper} queue system is used to read
  and preprocess files for the DNN training. A schematic overview of
  the pipeline used to preprocess batches of jets for the class
  prediction is given in Fig.~\ref{fig:tagger_queues}, while a similar
  queue is also used for the domain prediction. At the beginning of
  each epoch, a queue holding a randomised list of the input file
  names is initialised.  File names are dequeued asynchronously in
  multiple threads.  For each thread, \ROOT
  v6.18.00~\cite{Brun:1997pa} trees contained in the files are read
  from disk to memory in batches using a \tensorflow operation kernel,
  developed in the context of this paper. The resulting batches are
  resampled to achieve the same distributions in \pt and $\eta$ for
  all jet classes and are enqueued asynchronously into a second queue,
  which caches a list of tensors.  The DNN training commences by
  dequeuing a randomised batch of tensors and generating \ctau values
  for all SM jets within the batch.  The advantages of this system lie
  in its flexibility to adapt to new input features or samples
  on-the-fly. The reading of \ROOT trees and the (\pt, $\eta$)
  resampling for the SM jets proceeds asynchronously in multiple
  threads, managed by \tensorflow, on the CPU while the network is
  being trained.  A demonstration of the workflow can be found at
  Ref.~\cite{llpdnnx}.  }

\section{Validation with \texorpdfstring{$\Pp\Pp$}{pp} collision data}
\label{sec:validation}

\begin{figure*}[!th]
  \centering
  \includegraphics[width=0.48\textwidth]{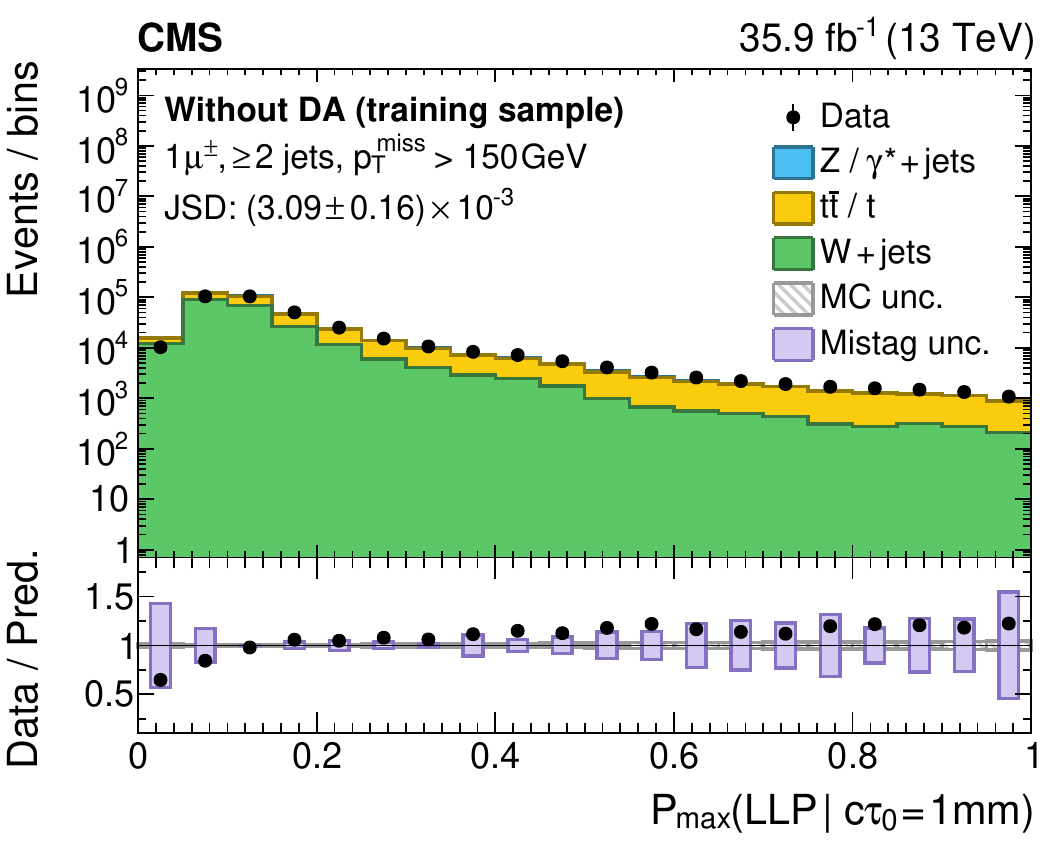}\hspace{0.02\textwidth}
  \includegraphics[width=0.48\textwidth]{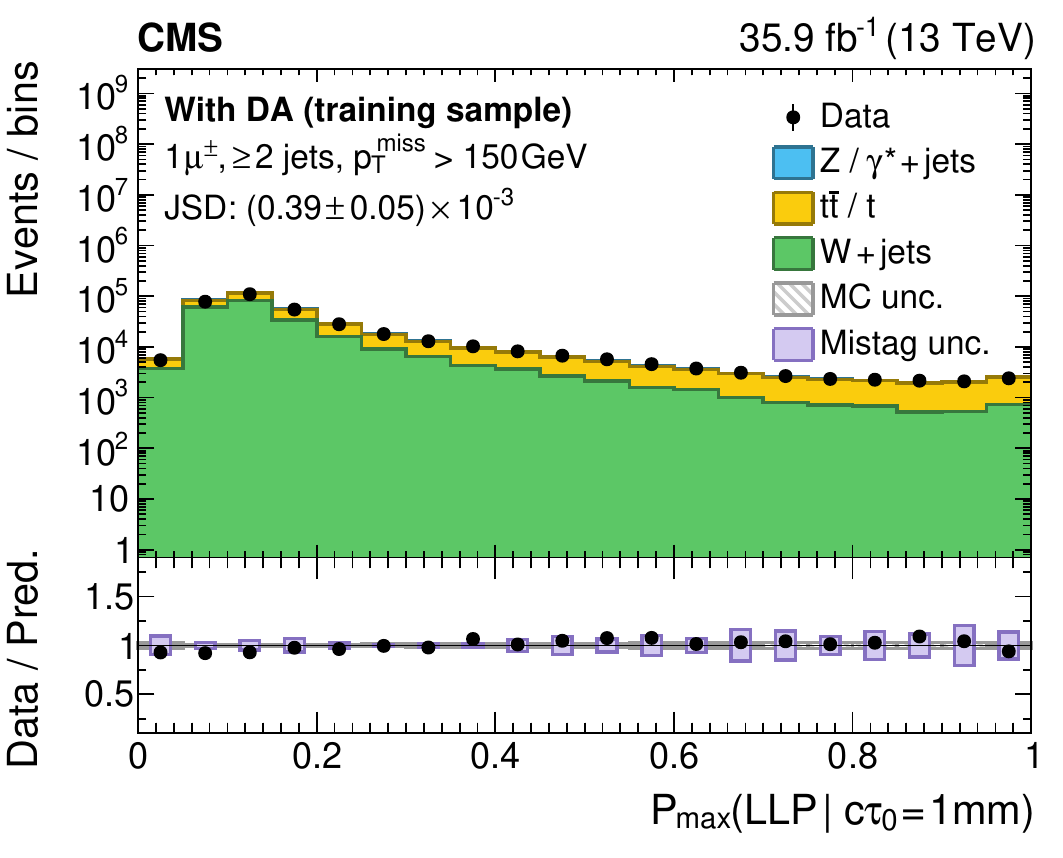} \\
  \includegraphics[width=0.48\textwidth]{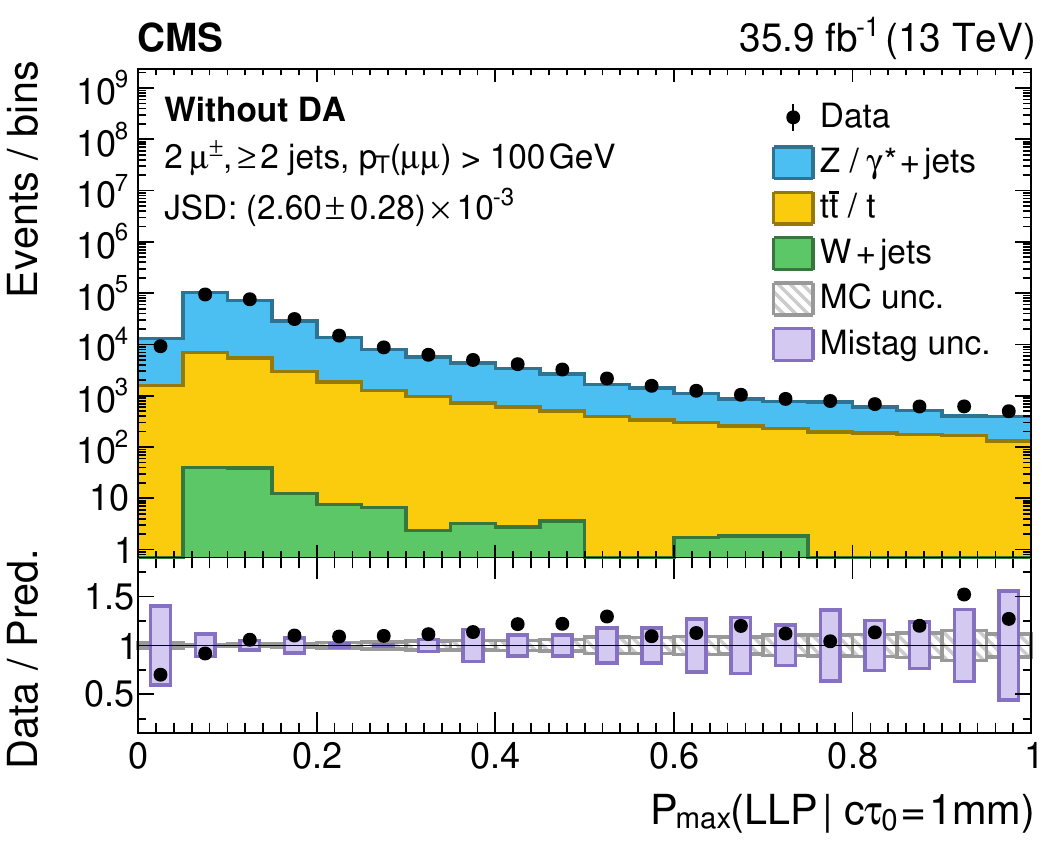}\hspace{0.02\textwidth}
  \includegraphics[width=0.48\textwidth]{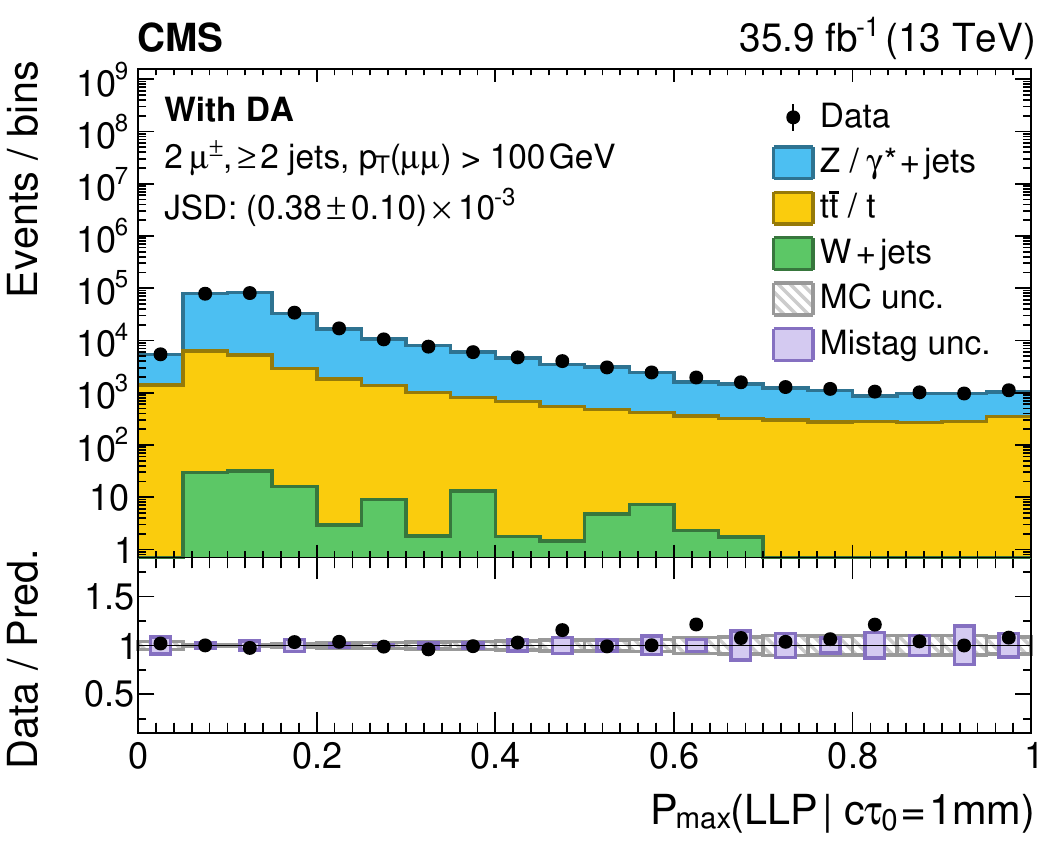} \\
  \caption{Distributions of the maximum probability for the LLP jet
    class obtained from all selected jets in a given event, \pmax. The
    distributions from $\Pp\Pp$ collision data (circular marker) and
    simulated events (histograms) are compared in the \mjets (upper
    row) and \mmjets (lower row) CRs, using a DNN trained without
    (left column) and with (right column) DA. All probabilities are
    evaluated with $\ctau = 1\mm$. The Jensen--Shannon divergence
    (JSD) is introduced in the text. The lower subpanels show the
    ratios of the binned yields obtained from data and Monte Carlo
    (MC) simulation. The statistical (hatched bands) and systematic
    (solid bands) uncertainties due to the finite-size simulation
    samples and the simulation mismodelling of the mistag rate,
    respectively, are also shown.\label{fig:tagger-validation} }
\end{figure*}

{In the absence of DA, the LLP jet probability \prob obtained from the
  simulation of the relevant SM backgrounds and the CR data can differ
  significantly, with deviations of up to 50\% in the binned counts of
  \prob distributions. A similar event-level variable is \pmax, which
  is defined as the maximum value of \prob obtained from all selected
  jets in a given event. A comparison of the \pmax distributions
  obtained from $\Pp\Pp$ collision data and simulated events in both
  the \mjets and \mmjets CRs, using a DNN trained with and without DA,
  is shown in Fig.~\ref{fig:tagger-validation}. The use of DA in the
  network leads to a significant improvement in the level of agreement
  in the binned counts of \pmax for the two domains of data and
  simulation, with only small residual differences remaining. This
  improvement is expected for the \mjets CR, as the same events are
  used to train, evaluate, and optimise the domain branch of the
  DNN. The \mmjets CR, comprising a statistically independent event
  sample, validates the method.  }

An estimate of the uncertainty in \pmax due to simulation mismodelling
is determined from jets in a statistically independent sample of
\mmjets events that satisfy $\pt(\PGm\PGm) < 100\GeV$. The magnitude
of the uncertainty is assessed by weighting up or down the simulated
events by the factor $w^{\pm} =
\prod^\text{jets}_{i}\Big(1\pm\big(\xi_{i}(\mathrm{LLP})-1\big)\Big)$,
where $\xi_{i}(\mathrm{LLP})$ is the ratio of counts from data and
simulation in bin $i$ of the \prob distribution.  The ratios of event
counts binned according to \pmax from pp collision data and
simulation, as well as the corresponding uncertainty estimates, are
shown in the lower panels of Fig.~\ref{fig:tagger-validation}. The
ratios are closer to unity, with reduced uncertainties, following the
application of DA.  The level of agreement between data and simulation
is further quantified by the Jensen--Shannon divergence
(JSD)~\cite{Lin:2006}, a measure of similarity between two probability
distributions that is bound to $[0,1]$ and takes a value of zero for
identical distributions. The JSD is reduced by an order of magnitude
following the application of DA. The quoted uncertainties in JSD
reflect the finite sizes of the data and simulated samples.

The application of DA leads to significantly reduced biases and
uncertainties in the modelling of \prob and related variables in the
signal-depleted CRs. This behaviour would translate into an improved
treatment for the estimation of SM backgrounds in the SR and the
associated experimental systematic uncertainties. However, only modest
gains in sensitivity to new high-mass particle states may be expected,
as the dominant uncertainties arise from the finite-size samples of pp
collision data and simulated events.

\section{Performance}
\label{sec:performance}

The performance of the tagger is demonstrated using simulated event
samples for split SUSY benchmark models with an uncompressed mass
spectrum, as defined in Section~\ref{sec:simulation}, and \ctau values
of 1\mm and 1\unit{m}. The two values of \ctau give greater weight to
the roles of the tracker and calorimeter systems,
respectively. Negligible SM background contributions are expected for
the 1\unit{m} scenario. An inclusive sample of \ttbar events is used
to provide both light-flavour ({\PQu}{\PQd}{\PQs}{\Pg}) jets, through
initial-state radiation and hadronic decays of the {\PW} boson, and
{\PQb} jets, with $\pt > 30\GeV$ and $\abs{\eta} < 2.4$.

\begin{figure}[!th]
  \centering
  \includegraphics[width=0.48\textwidth]{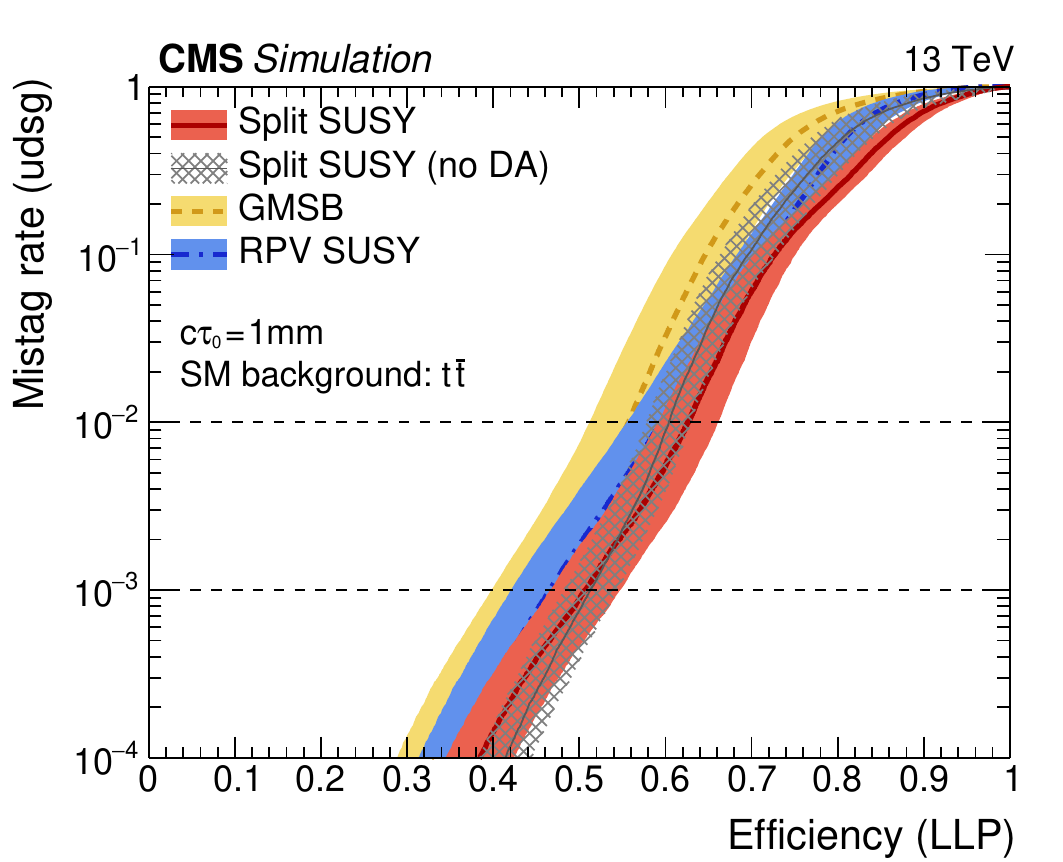}\hspace{0.03\textwidth}
  \includegraphics[width=0.48\textwidth]{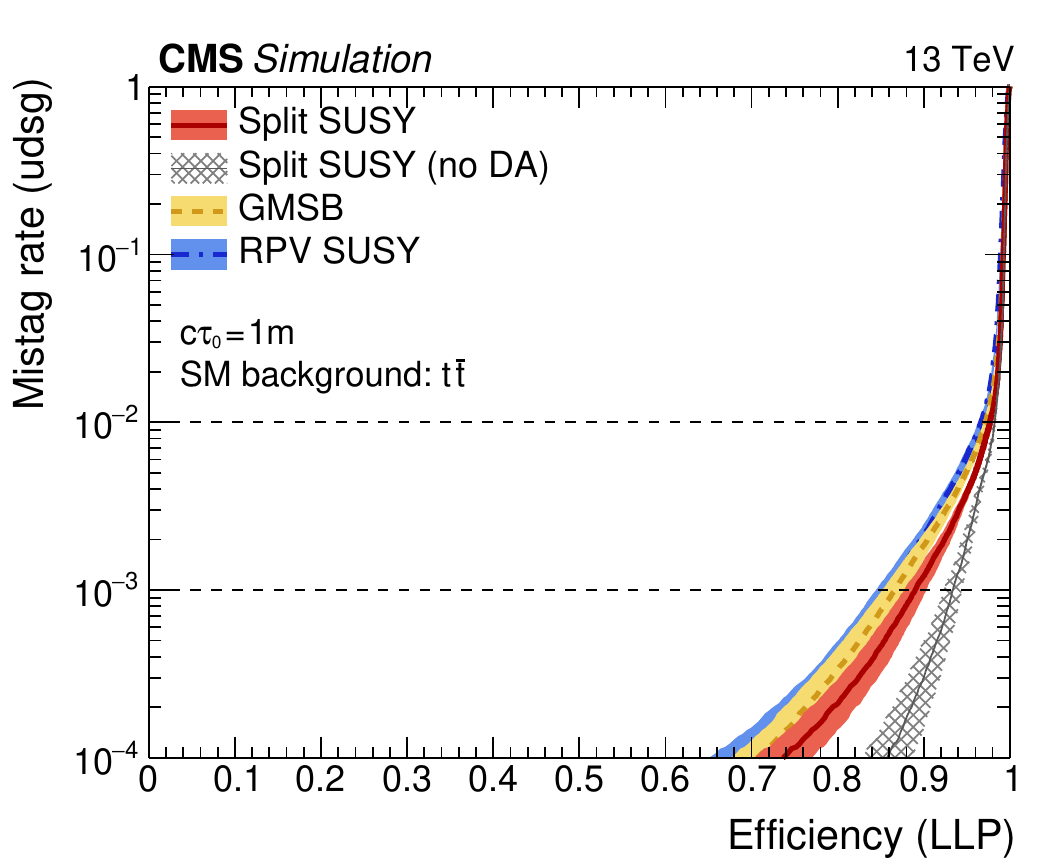}
  \caption{The ROC curves illustrating the tagger performance for the
    split (solid line), GMSB (dashed line), and RPV (dot-dashed line)
    SUSY benchmark models, defined in Section~\ref{sec:simulation},
    assuming \ctau values of 1\mm (\cmsLeft) and 1\unit{m}
    (\cmsRight). The thin line with hatched shading indicates the
    performance obtained with a DNN training using split SUSY samples
    but without the DA. The jet sample is defined in the
    text. \label{fig:roc}}
\end{figure}

The efficiency of the tagger to identify correctly the LLP jet class
depends on the chosen working point, defined by a threshold
requirement on the jet class probability. The mistag rates for the
remaining jet classes also depend on the same working point. The
receiver operating characteristic (ROC) curves that provide the LLP
jet tagging efficiency and the mistag rate for the
{\PQu}{\PQd}{\PQs}{\Pg} jet class as a function of the working point
are shown in Fig.~\ref{fig:roc}. The uncertainties indicated by the
shaded bands are determined from the standard deviation obtained from
a ten-fold cross validation. Given a mistag rate of 0.01\%, equivalent
to a background rejection factor of 10\,000, efficiencies of 40 and
70\% are obtained for split SUSY models with \ctau values of 1\mm and
1\unit{m}, respectively. These efficiencies decrease by a factor
${\approx}0.6$ if the DNN relies solely on the global jet
features. Additional studies reveal that the tagger efficiency for the
LLP jet class falls to 25\% in order to achieve the same mistag rate
of 0.01\% for the (SM) \PQb jet class when considering split SUSY
models with $\ctau = 1\mm$.

Figure~\ref{fig:roc} also shows ROC curves when evaluating the DNN
using simulated events from the GMSB and RPV SUSY benchmark models, as
defined in Section~\ref{sec:simulation}. The jets originate from
{\PQu}{\PQd}{\PQs} quarks (gluons) from the gluino decay in the case
of split (GMSB) SUSY models, and {\PQb} hadrons from the top squark
decay for RPV SUSY. A similar level of performance is observed for
these SUSY models, in which the LLP jets have a different underlying
flavour. Furthermore, the ROC curves indicated by the thick and thin
solid curves illustrate the tagger performance when the DNN is trained
with or without DA, respectively, for the split SUSY benchmark
models. Studies demonstrate a comparable performance for split SUSY
models with and without DA for the region $\ctau \leq 10\mm$. For
larger values of \ctau, the performance is overestimated in the
absence of DA, as indicated by the $\ctau = 1\unit{m}$ scenario shown
in Fig.~\ref{fig:roc}. This is because the DNN is able to exploit
patterns in the features obtained from simulation that are not
representative of those obtained from data.

\begin{figure*}[!th]
  \centering
  \includegraphics[width=0.99\textwidth]{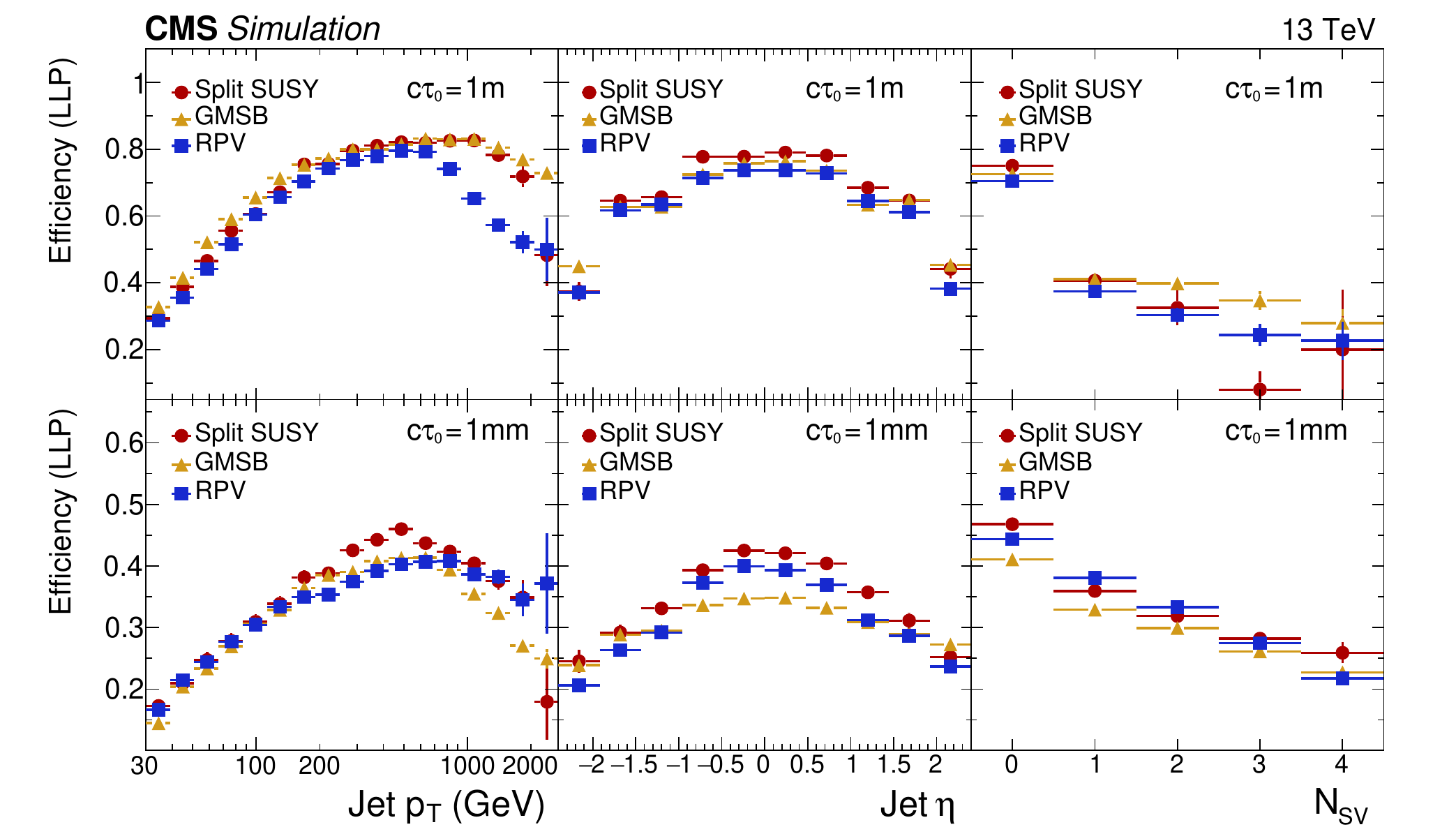}
  \caption{The LLP jet tagging efficiency as a function of the jet
    \pt, $\eta$, and $N_{\mathrm{SV}}$ using a working point that
    yields a mistag rate of 0.01\% for the {\PQu}{\PQd}{\PQs}{\Pg} jet
    class, as obtained from an inclusive sample of simulated \ttbar
    events. The efficiency curves are shown separately for the split
    (circular marker), GMSB (triangle marker), and RPV (square marker)
    SUSY benchmark models, defined in Section~\ref{sec:simulation},
    assuming \ctau values of 1\unit{m} (upper row) and 1\mm (lower
    row). \label{fig:projection}}
\end{figure*}

The LLP jet tagging efficiency is shown in Fig.~\ref{fig:projection}
as a function of the jet~\pt, $\eta$, and the number of reconstructed
secondary vertices within the jet (\nsv) for a working point that
yields a mistag rate of 0.01\% for the {\PQu}{\PQd}{\PQs}{\Pg} jet
class obtained from simulated \ttbar events. Efficiencies are highest
for high-\pt, centrally produced jets with $\nsv = 0$. The latter
observation demonstrates the complementary performance of the tagger
with respect to a more standard approach of relying on reconstructed
secondary vertices to identify displaced jets.

\begin{figure}[!th]
  \centering
  \includegraphics[width=0.48\textwidth]{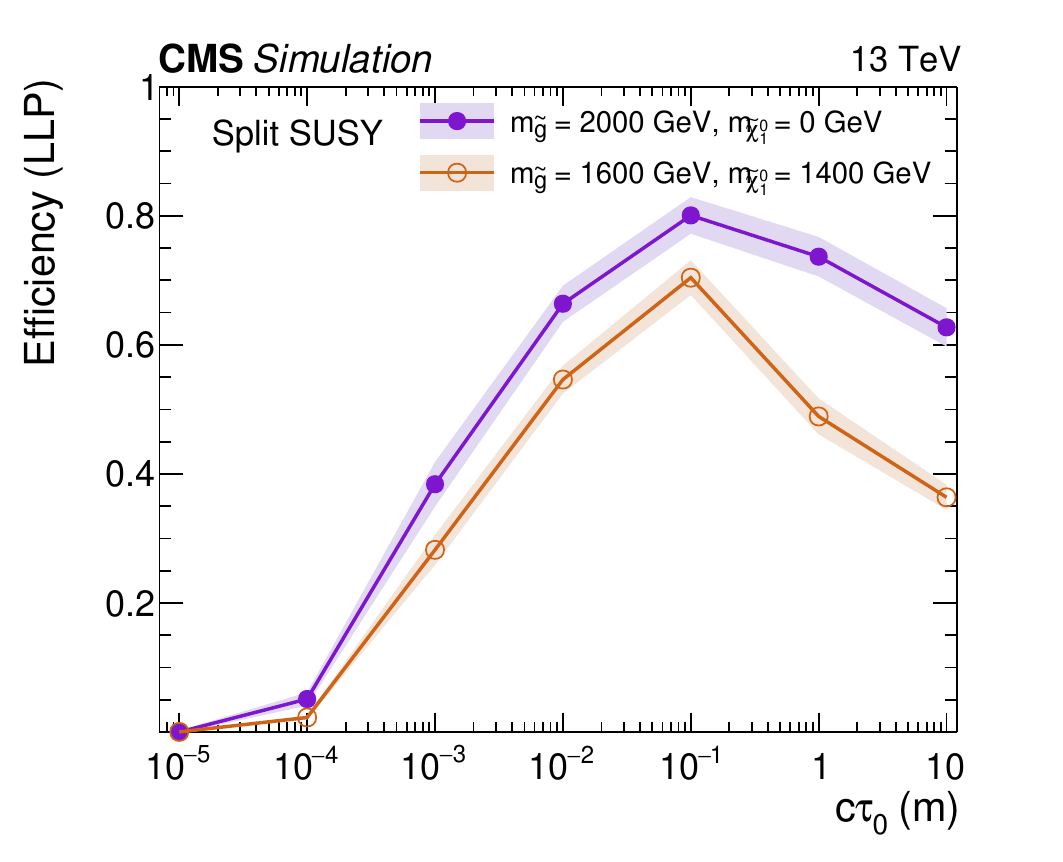}\hspace{0.03\textwidth}
  \includegraphics[width=0.48\textwidth]{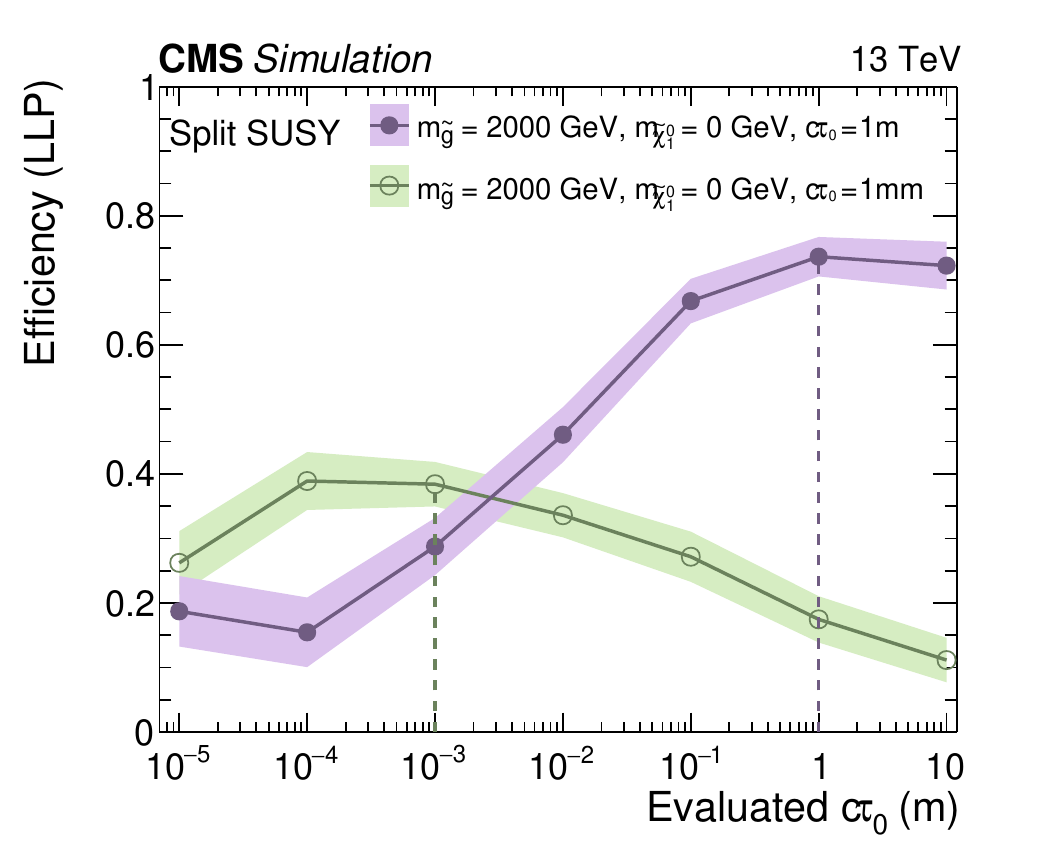}
  \caption{The LLP jet tagging efficiency, using a working point that
    yields a mistag rate of 0.01\% for the {\PQu}{\PQd}{\PQs}{\Pg}
    jet class obtained from an inclusive sample of simulated \ttbar
    events, when (\cmsLeft) the DNN is evaluated as a function of the
    model parameter value \ctau for an uncompressed and a compressed
    split SUSY model, and (\cmsRight) the DNN is evaluated over a
    range of \ctau values for uncompressed split SUSY models generated
    with $\ctau = 1\mm$ and 1\unit{m}; the dashed vertical lines
    indicate equality for the evaluated and generated values of \ctau
    for each model. The fixed model parameters are defined in the
    legends.
    \label{fig:wrong}}
\end{figure}

The performance of the DNN parameterisation according to \ctau is
shown in Figure~\ref{fig:wrong}. The \cmsLeft panel shows the LLP jet
tagging efficiency, using a working point that yields a mistag rate of
0.01\% for the {\PQu}{\PQd}{\PQs}{\Pg} jet class, as a function of the
generated \ctau value, for both an uncompressed and a compressed split
SUSY model. Efficiencies in the range 40--80 (30--70)\% are achieved
for uncompressed (compressed) scenarios with $1\mm \leq \ctau \leq
10\unit{m}$. The compressed scenarios are characterised by low-\pt
jets, resulting in lower efficiencies as shown in Fig.~\ref{fig:projection}.

The performance is further tested using two uncompressed split SUSY
models with $\ctau = 1\mm$ and 1\unit{m}. The LLP jet tagging
efficiency is obtained by evaluating the DNN for each value of \ctau
in the range $10\mum \leq \ctau \leq 10\unit{m}$. Again, the
efficiency is determined using a working point that is tuned for each
evaluated \ctau value to yield a mistag rate of 0.01\% for the
{\PQu}{\PQd}{\PQs}{\Pg} jet class. The efficiency as a function of the
evaluated \ctau value is shown in Fig.~\ref{fig:wrong} (\cmsRight). The
maximum efficiency is obtained when the evaluated value of \ctau
approximately matches the parameter value of the split SUSY
model. This behaviour may be used to characterise a potential signal
contribution in terms of the model parameter \ctau. Finally, studies
demonstrate that the parameterised approach does not significantly
impact performance with respect to the training of multiple DNNs, one
per \ctau value.

\section{Application to a search for LLPs}
\label{sec:showcase}

The performance of the LLP jet tagger is demonstrated by applying it
to the search for long-lived gluinos with $10\mum \leq \ctau \leq
10\unit{m}$.  Expected limits on the theoretical production cross
section for gluino pairs are determined. The search is performed using
statistically independent samples of simulated events to ensure an
unbiased evaluation of the tagger.

\subsection{Categorisation of events and uncertainties}
\label{sec:result}

Candidate events that satisfy the SR requirements defined in
Section~\ref{sec:selections} are categorised according to: the number
of jets in the event, \njet; the number of jets for which \prob is
above a predefined threshold, \ntag; and \HT. The resulting six
categories are summarised in Table~\ref{tab:showcase-binning}. Models
with an uncompressed mass spectrum, $\mgluino - \mlsp \gtrsim
200\GeV$, are characterised by high values of \HT. Models with a
compressed mass spectrum, $\mgluino - \mlsp \lesssim 200\GeV$, are
characterised by lower values of \HT because of the limited kinematic
phase space available to the jets from the gluino decay and an
increased reliance on associated jet production from initial-state
radiation. Events that satisfy $\ntag < 2$ are grouped into a single
additional category, which is used to constrain the normalisation of
simulated background events during the statistical evaluation.

\begin{table*}[!th]
  \centering
  \topcaption{\label{tab:showcase-binning}
    The event counts and uncertainties for SM backgrounds and split
    SUSY models, as determined from simulation, in categories defined
    by \HT and (\njet, \ntag). The simulated samples are normalised
    to an integrated luminosity of 35.9\fbinv. The uncompressed and
    compressed split SUSY  models are defined in
    Section~\ref{sec:simulation}. The value of \ctau is assumed to be
    1\mm. The uncertainties include both statistical and
    systematic contributions. Expected counts for events that satisfy
    $\ntag < 2$ are not shown.
  }
  \renewcommand{\arraystretch}{1.3}
  \begin{tabular}{@{}l r@{$\:\pm\:$}l  r@{$\:\pm\:$}l  r@{$\:\pm\:$}l @{}l r@{$\:\pm\:$}l  r@{$\:\pm\:$}l  r@{$\:\pm\:$}l@{}}
    \hline
    \HT\,($\GeVns$) & \multicolumn{2}{c}{300--800}
                    & \multicolumn{2}{c}{300--800}
                    & \multicolumn{2}{c}{300--800}
                    &  & \multicolumn{2}{c}{$>$800}
                    & \multicolumn{2}{c}{$>$800}
                    & \multicolumn{2}{c@{}}{$>$800} \\
    (\njet, \ntag)   & \multicolumn{2}{c}{(3--4, $\geq$2)}
                     & \multicolumn{2}{c}{(5, $\geq$2)}
                     & \multicolumn{2}{c}{($\geq$6, $\geq$3)}
                     &  & \multicolumn{2}{c}{(3--4, $\geq$2)}
                     & \multicolumn{2}{c}{(5, $\geq$2)}
                     & \multicolumn{2}{c@{}}{($\geq$6, $\geq$3)} \\
    \hline
    \znunujets       & 41  & 39  & 6.5  & 5.8 & 0.6 & 0.4 &  & 3.3 & 2.8 & 1.6 & 1.2 & 0.1 & 0.1                  \\
    \wlnujets        & 56  & 44  & 11.6 & 5.1 & 1.5 & 0.5 &  & 3.6 & 2.5 & 1.2 & 3.0 & \multicolumn{2}{c}{$<$0.1} \\
    \ttbar           & 40  & 36  & 18   & 16  & 1.9 & 1.1 &  & 2.1 & 1.3 & 3.2 & 2.4 & 3.0 & 2.1                  \\
    Single top quark & 5.7 & 5.2 & 2.6  & 2.2 & 0.3 & 0.2 &  & 0.6 & 0.4 & 0.5 & 0.3 & 0.4 & 0.3                  \\[\cmsTabSkip]
    Total SM         & 142 & 69  & 39   & 18  & 4.3 & 1.3 &  & 9.7 & 4.0 & 6.5 & 4.1 & 3.5 & 2.5                  \\
    Uncompressed & \multicolumn{2}{c}{$<$0.1}
                 & \multicolumn{2}{c}{$<$0.1}
                 & \multicolumn{2}{c}{$<$0.1}
                 &  & 3.0 & 2.9 & 3.8 & 3.7 & 5.7 & 5.5 \\
    Compressed & 5.4 & 5.0 & 4.2 & 3.8 & 2.8 & 2.5 &  & 1.1 & 0.9 & 2.5 & 2.2 & 4.5 & 4.1 \\
    \hline
\end{tabular}
\end{table*}

The tagger is evaluated using simulated samples of all relevant SM
backgrounds, described in Section~\ref{sec:simulation}. The negligible
background contribution from multijet events in the SR is not
considered in this exploratory study.  The predefined threshold on
\prob is determined per \ctau value such that the most sensitive event
categories are nearly free of SM background contributions, while
control over uncertainties due to the finite-size simulated samples is
maintained. The \prob thresholds fall in the range 30--50\% and yield
an LLP jet tagging efficiency of about 30--90\% for $\ctau \geq
1\mm$. Table~\ref{tab:showcase-binning} summarises the expected counts
and uncertainties for the contributions from SM backgrounds, in the
various event categories, for $\ctau = 1\mm$. The statistical
uncertainty arising from the finite size of the simulated samples is
the dominant contribution to the quoted uncertainties. Additional
systematic contributions are described below. The expected event
counts from the uncompressed and compressed benchmark models of split
SUSY, defined in Section~\ref{sec:simulation}, are also provided.

Several sources of systematic uncertainty in the SM background
expectations are considered. An uncertainty of ${\pm}20\%$ is assumed
in the normalisation of each dominant background process, \wlnujets,
\znunujets, and single top quark and \ttbar production, which is
motivated by theoretical uncertainties in the production cross
sections and uncertainties in the experimental acceptance for the
final-state leptons~\cite{Kallweit:2015dum}. The uncertainty in the
mistag rate for jets from SM processes is typically below 10\%,
determined by the procedure described in
Section~\ref{sec:validation}. The jet energy is varied within its
uncertainty and resolution~\cite{Khachatryan:2016kdb}, and the
resulting shifts are propagated to \ptvecmiss. The unclustered
component of \ptvecmiss is varied within its uncertainties. The
uncertainty in the number of pileup interactions is determined by
varying the inelastic $\Pp\Pp$ cross section within its uncertainty of
$\pm$5\%~\cite{Sirunyan:2018nqx}. The renormalisation and
factorisation scales of the aforementioned four dominant SM
backgrounds are varied independently per process by factors of 0.5 and
2~\cite{Kalogeropoulos:2018cke} to estimate the migration of events
between categories. An uncertainty of ${\pm}2.5$\% in the integrated
luminosity is assumed~\cite{CMS-PAS-LUM-17-001}.

The LLP jet tagging efficiency determined from simulated events of
split SUSY models, \emc, may require the application of corrections to
account for sources of potential mismodelling. The corrections are
applied by reweighting simulated events as follows:
\begin{linenomath}
\ifthenelse{\boolean{cms@external}}
{
\begin{equation}
  \begin{aligned}
  w &= \left(\frac{1-\edata}{1-\emc}\right)^{(\nllp-\ntllp)}
      \,
      \left(\frac{\edata}{\emc}
      \right)^{\ntllp}\\
    &=
    \left(
      \frac{1-\sfac\,\emc}{1-\emc}
    \right)^{(\nllp-\ntllp)}
    \,
    \sfac^{\ntllp},
  \end{aligned}
\end{equation}
}
{
\begin{equation}
w = \left(\frac{1-\edata}{1-\emc}\right)^{(\nllp-\ntllp)}
      \,
      \left(\frac{\edata}{\emc}
      \right)^{\ntllp}
    =
    \left(
      \frac{1-\sfac\,\emc}{1-\emc}
    \right)^{(\nllp-\ntllp)}
    \,
    \sfac^{\ntllp},
  \end{equation}
}
\end{linenomath}
where \nllp denotes the number of LLP jets and \ntllp the number of
LLP jets that are also tagged per event.  The scale factor \sfac=
\edata/\emc denotes a correction to \emc, and the product of \emc and
the \sfac is bound to $[0,1]$.

\subsection{Signal hypothesis testing}
\label{sec:signal-eff}
A likelihood model is used to test for the presence of new-physics
signals in the SR. The observed event count in each event category is
modelled as a Poisson-distributed variable around the sum of the SM
expectation and a potential signal contribution. The expected event
counts from SM backgrounds are obtained from the simulated
samples. The uncertainties resulting from the finite simulated samples
are modelled using the Barlow--Beeston
method~\cite{Barlow:1993dm}. The systematic uncertainties in the SM
background estimates are accommodated in the likelihood model as
nuisance parameters.

Hypothesis testing is performed using an Asimov data
set~\cite{Cowan:2010js} to provide expected constraints on simplified
models of split SUSY. A modified frequentist approach is used to
determine the expected upper limits at 95\% confidence level (\CL) on
the theoretical gluino pair production cross section as a function of
\mgluino, \mlsp, and \ctau. The signal strength parameter,
$r_\text{UL}$, expresses the upper limit on the production cross
section relative to the theoretical value. Alternatively, expected
lower limits at 95\% \CL on \mgluino can be determined as a function
of \ctau. The approach is based on the profile likelihood ratio as the
test statistic~\cite{CMS-NOTE-2011-005}, the \CLs
criterion~\cite{junk, read}, and the asymptotic
formulae~\cite{Cowan:2010js} to approximate the distributions of the
test statistic under the background-only and signal-plus-background
hypotheses.

\subsection{Interpretations}
\label{sec:limits}

\begin{figure}[!th]
  \centering
  \includegraphics[width=0.48\textwidth]{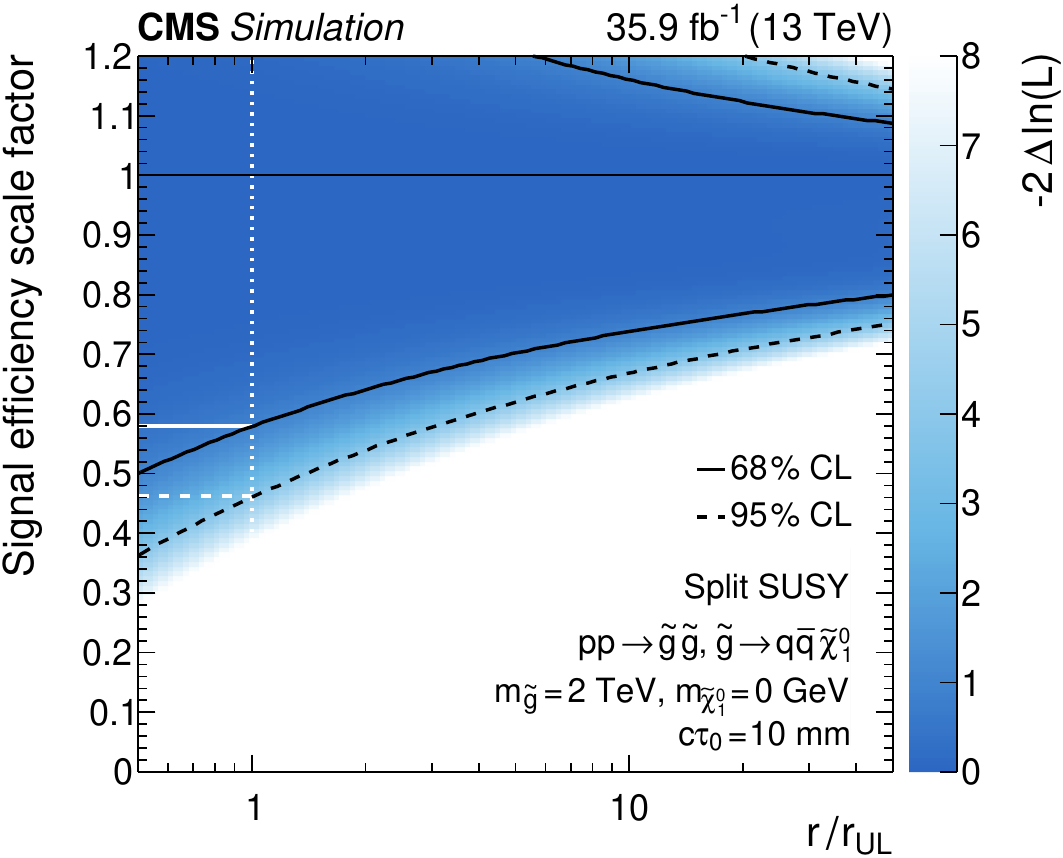}\hspace{0.03\textwidth}
  \includegraphics[width=0.48\textwidth]{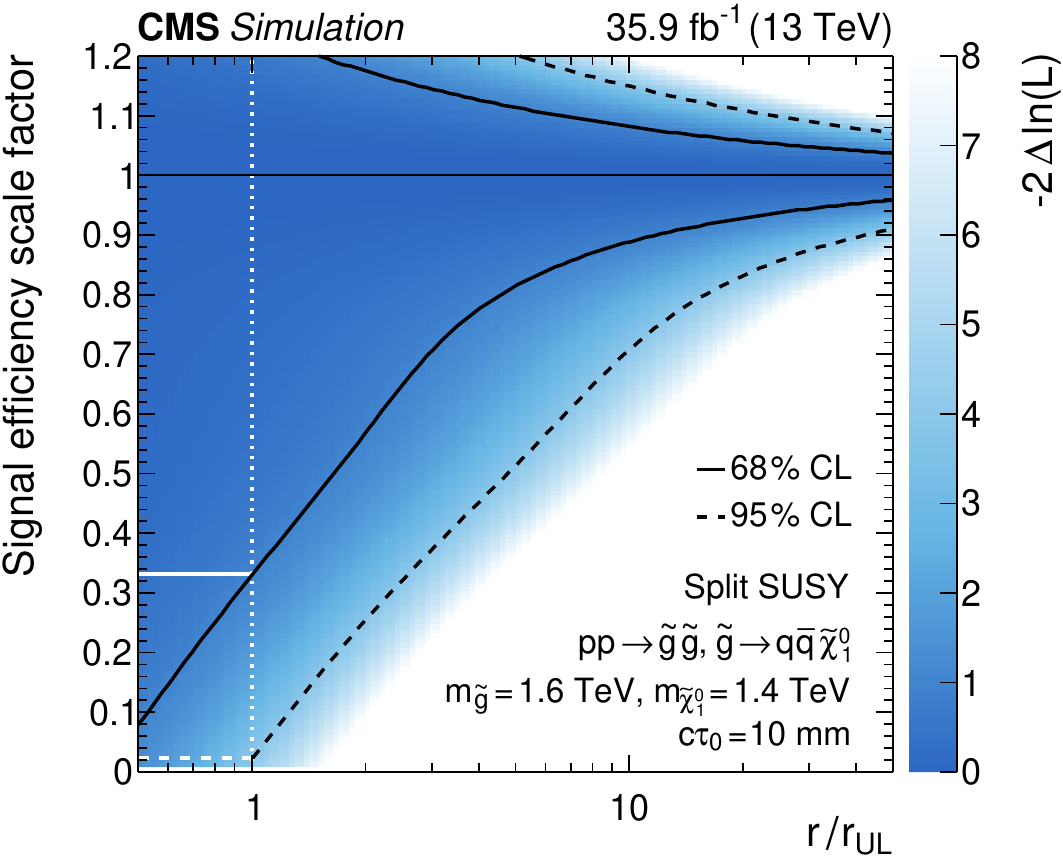}
  \caption{\label{fig:showcase-llpeff}The negative log-likelihood of a
    maximum likelihood fit to the Asimov data set as a function of the
    signal efficiency scale factor and $r/r_\text{UL}$ for a
    (\cmsLeft) uncompressed and (\cmsRight) compressed scenario. The
    black solid (dashed) line indicate the 68 (95)\% \CL interval,
    while for $r = r_\text{UL}$ (white dotted line) the white solid
    and dashed lines indicate the \sfac constraints at 68\% and 95\%
    \CL, respectively.  The product of the LLP jet tagger efficiency
    and the \sfac is bound to $[0,1]$.}
\end{figure}

Figure~\ref{fig:showcase-llpeff} shows the negative log-likelihood
from a maximum likelihood fit to the Asimov data as a function of both
the \sfac and $r/r_\text{UL}$, where the signal strength parameter $r$
represents the injected gluino pair production cross section relative
to the theoretical value, for two split SUSY benchmark scenarios with
an uncompressed and compressed mass spectrum, and $\ctau = 10\mm$. The
model assumptions are indicated in the figure legends. The \sfac is
constrained to $>$0.6 ($>$0.3) at 68\% \CL for $r/r_\text{UL} = 1$ for
the uncompressed (compressed) model, and it is bound to $\lesssim$1.2
by the condition $\sfac\,\emc \in [0,1]$.  The figure demonstrates
that the likelihood is not degenerate with respect to the scale factor
and signal strength. Instead, the (\njet, \ntag) categorisation scheme
allows the SF to be constrained \textit{in situ} in the
signal-plus-background fit when adding the SF as a nuisance parameter
to the likelihood model, described in Section~\ref{sec:result}.

\begin{figure}[!th]
  \centering
  \includegraphics[width=0.48\textwidth]{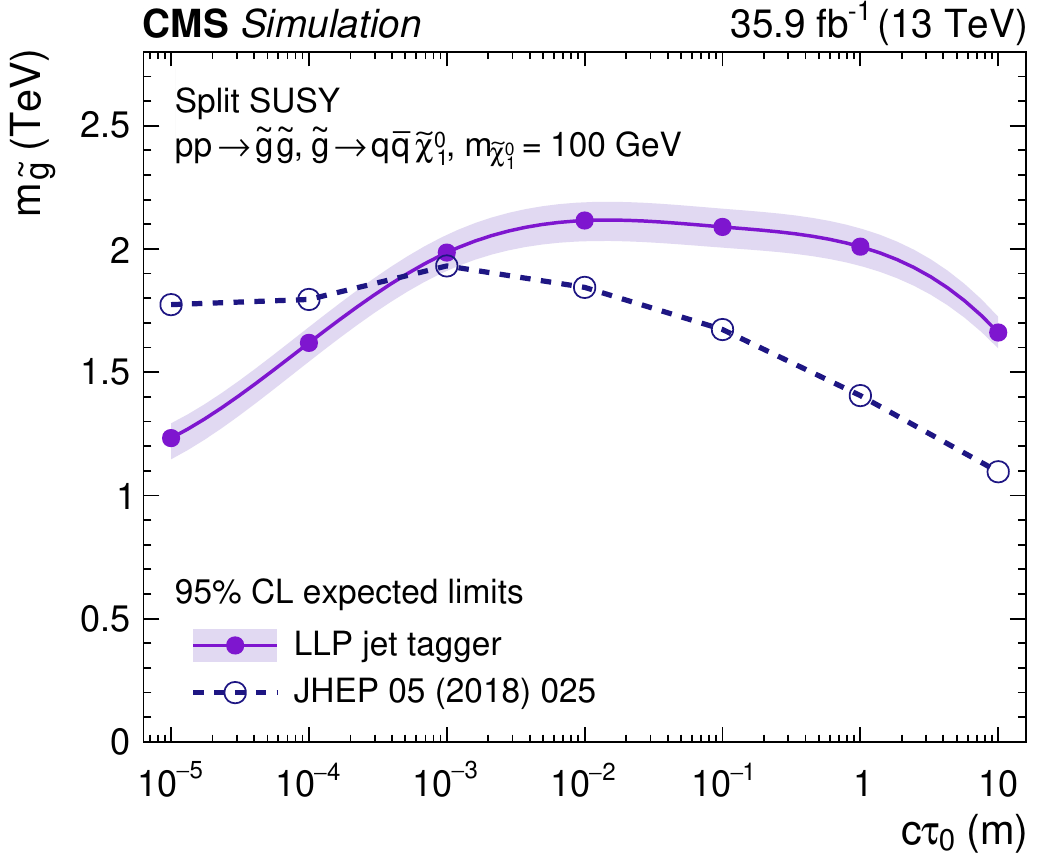}\hspace{0.03\textwidth}
  \includegraphics[width=0.48\textwidth]{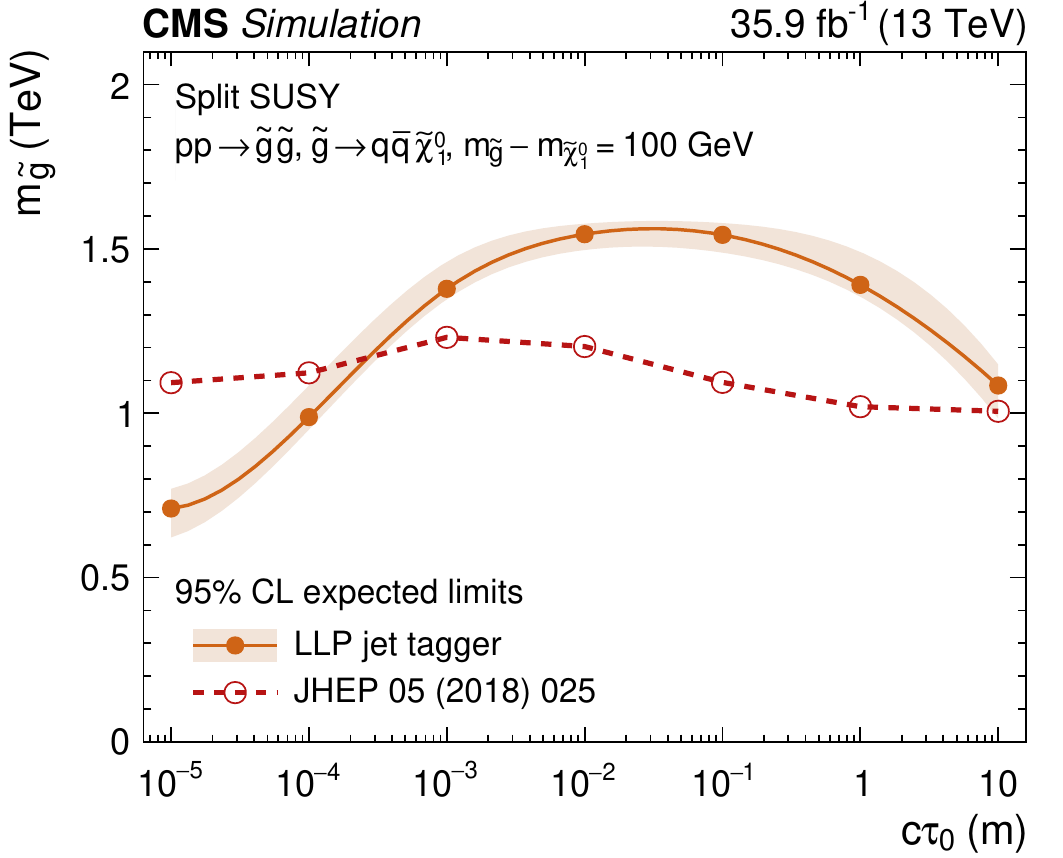}
  \caption{Expected 95\% \CL lower limits on \mgluino as a function of
    \ctau for split SUSY models with an uncompressed (\cmsLeft) and a
    very compressed (\cmsRight) mass spectrum. The shaded bands
    indicate the total uncertainty from both statistical and
    systematic sources. The model assumptions are indicated by the
    legends. The results are compared to the expected limits obtained
    in Ref.~\cite{Sirunyan:2018vjp}, indicated by the dashed
    lines. \label{fig:showcase-summary}}
\end{figure}

Figure~\ref{fig:showcase-summary} summarises the expected lower limit
on \mgluino (95\% \CL) as a function of \ctau for simplified models of
split SUSY that assume the production of gluino pairs. The model
assumptions are indicated by the legend in each panel. The \cmsLeft
(\cmsRight) panel presents the expected mass exclusions for models with an
uncompressed (compressed) mass spectrum. A lower limit on the gluino
mass in excess of 2.1\,(1.5)\TeV is obtained for models with an
uncompressed (compressed) mass spectrum. The mass exclusions are
determined by assuming an integrated luminosity of 35.9\fbinv. This
permits a comparison with the exclusions reported by an inclusive
search for SUSY~\cite{Sirunyan:2018vjp} in final states containing
jets and \ptmiss, over the same range in \ctau values. Significant
gains in excluded values of \mgluino, of up to approximately
$500\GeV$, are expected for $\ctau \gtrsim 1\mm$. The coverage is also
competitive with respect to a dedicated reconstruction technique that
is reported in Ref.~\cite{Aaboud:2017iio}. For the region $\ctau <
1\mm$, the tagger performance degrades because of a limited ability to
tag an LLP jet in the vicinity of the primary $\Pp\Pp$ interaction
vertex, while the reference search is able to exploit the
distinguishing kinematical features of split SUSY events through a
finer categorisation of candidate signal events. For the region $\ctau
> 1\unit{m}$, the LLPs frequently decay outside the experimental
acceptance, which leads to an increased reliance on the presence of
initial-state radiation.

\newpage

\section{Summary}
\label{sec:summary}

Many models of new physics beyond the standard model predict the
production of long-lived particles (LLPs) in proton-proton ($\Pp\Pp$)
collisions at the LHC. Jets arising from the decay of LLPs (LLP jets)
can be appreciably displaced from the $\Pp\Pp$ collisions. A novel
tagger to identify LLP jets is presented. The tagger employs a deep
neutral network (DNN) using an architecture inspired by the CMS
DeepJet algorithm. Simplified models of split supersymmetry (SUSY),
which yield neutralinos and LLP jets from the decay of long-lived
gluinos, are used to train the DNN and demonstrate its performance.

The application of various techniques related to the tagger are
reported. A custom labelling scheme for LLP jets based on
generator-level information from Monte Carlo programs is defined. The
proper decay length \ctau of the gluino is used as an external
parameter to the DNN, which allows hypothesis testing over several
orders of magnitude in \ctau with a single DNN. The DNN is trained
using samples of both simulated events and $\Pp\Pp$ collision
data. The application of domain adaptation by backward propagation
significantly improves the agreement of the DNN output for simulation
and data, by an order of magnitude according to the Jensen--Shannon
divergence, when compared to training the DNN with simulation
only. The method is validated using signal-depleted control samples of
$\Pp\Pp$ collisions at a centre-of-mass energy of 13\TeV. The samples
were recorded by the CMS experiment and correspond to an integrated
luminosity of 35.9\fbinv. Training the DNN with $\Pp\Pp$ collision
data does not significantly degrade the tagger performance. The tagger
rejects 99.99\% of light-flavour jets from standard model processes,
as measured in an inclusive \ttbar sample, while retaining
approximately 30--80\% of LLP jets for split SUSY models with $1\mm
\leq \text{c}\tau_0 \leq 10\unit{m}$ and a gluino-neutralino mass
difference of at least 200\GeV.

Finally, the potential performance of the tagger is demonstrated in
the framework of a search for split SUSY in final states containing
jets and significant missing transverse momentum. Simulated event
samples provide the expected contributions from standard model
background processes. Candidate signal events were categorised
according to the scalar sum of jet momenta, the number of jets, and
the number of tagged LLP jets. Expected lower limits on the gluino
mass at 95\% confidence level are determined with a binned likelihood
fit as a function of \ctau in the range from 10\mum to 10\unit{m}. A
procedure to constrain a correction to the LLP jet tagger efficiency
in the likelihood fit is introduced. Competitive limits are
demonstrated: models with a long-lived gluino of mass
${\gtrsim}2\TeV$, a neutralino mass of 100\GeV, and a proper decay
length in the range $1\mm \leq \ctau \leq 1\unit{m}$ are expected to
be excluded by this search.

\begin{acknowledgments}
  We congratulate our colleagues in the CERN accelerator departments
  for the excellent performance of the LHC and thank the technical and
  administrative staffs at CERN and at other CMS institutes for their
  contributions to the success of the CMS effort. In addition, we
  gratefully acknowledge the computing centers and personnel of the
  Worldwide LHC Computing Grid for delivering so effectively the
  computing infrastructure essential to our analyses. Finally, we
  acknowledge the enduring support for the construction and operation
  of the LHC and the CMS detector provided by the following funding
  agencies: BMBWF and FWF (Austria); FNRS and FWO (Belgium); CNPq,
  CAPES, FAPERJ, FAPERGS, and FAPESP (Brazil); MES (Bulgaria); CERN;
  CAS, MoST, and NSFC (China); COLCIENCIAS (Colombia); MSES and CSF
  (Croatia); RPF (Cyprus); SENESCYT (Ecuador); MoER, ERC IUT, PUT and
  ERDF (Estonia); Academy of Finland, MEC, and HIP (Finland); CEA and
  CNRS/IN2P3 (France); BMBF, DFG, and HGF (Germany); GSRT (Greece);
  NKFIA (Hungary); DAE and DST (India); IPM (Iran); SFI (Ireland);
  INFN (Italy); MSIP and NRF (Republic of Korea); MES (Latvia); LAS
  (Lithuania); MOE and UM (Malaysia); BUAP, CINVESTAV, CONACYT, LNS,
  SEP, and UASLP-FAI (Mexico); MOS (Montenegro); MBIE (New Zealand);
  PAEC (Pakistan); MSHE and NSC (Poland); FCT (Portugal); JINR
  (Dubna); MON, RosAtom, RAS, RFBR, and NRC KI (Russia); MESTD
  (Serbia); SEIDI, CPAN, PCTI, and FEDER (Spain); MOSTR (Sri Lanka);
  Swiss Funding Agencies (Switzerland); MST (Taipei); ThEPCenter,
  IPST, STAR, and NSTDA (Thailand); TUBITAK and TAEK (Turkey); NASU
  (Ukraine); STFC (United Kingdom); DOE and NSF (USA).

  \hyphenation{Rachada-pisek} Individuals have received support from
  the Marie-Curie program and the European Research Council and
  Horizon 2020 Grant, contract Nos.\ 675440, 752730, and 765710
  (European Union); the Leventis Foundation; the A.P.\ Sloan
  Foundation; the Alexander von Humboldt Foundation; the Belgian
  Federal Science Policy Office; the Fonds pour la Formation \`a la
  Recherche dans l'Industrie et dans l'Agriculture (FRIA-Belgium); the
  Agentschap voor Innovatie door Wetenschap en Technologie
  (IWT-Belgium); the F.R.S.-FNRS and FWO (Belgium) under the
  ``Excellence of Science -- EOS'' -- be.h project n.\ 30820817; the
  Beijing Municipal Science \& Technology Commission,
  No. Z181100004218003; the Ministry of Education, Youth and Sports
  (MEYS) of the Czech Republic; the Deutsche Forschungsgemeinschaft
  (DFG) under Germany's Excellence Strategy -- EXC 2121 ``Quantum
  Universe'' -- 390833306; the Lend\"ulet (``Momentum") Program and
  the J\'anos Bolyai Research Scholarship of the Hungarian Academy of
  Sciences, the New National Excellence Program \'UNKP, the NKFIA
  research grants 123842, 123959, 124845, 124850, 125105, 128713,
  128786, and 129058 (Hungary); the Council of Science and Industrial
  Research, India; the HOMING PLUS program of the Foundation for
  Polish Science, cofinanced from European Union, Regional Development
  Fund, the Mobility Plus program of the Ministry of Science and
  Higher Education, the National Science Center (Poland), contracts
  Harmonia 2014/14/M/ST2/00428, Opus 2014/13/B/ST2/02543,
  2014/15/B/ST2/03998, and 2015/19/B/ST2/02861, Sonata-bis
  2012/07/E/ST2/01406; the National Priorities Research Program by
  Qatar National Research Fund; the Ministry of Science and Education,
  grant no. 14.W03.31.0026 (Russia); the Programa Estatal de Fomento
  de la Investigaci{\'o}n Cient{\'i}fica y T{\'e}cnica de Excelencia
  Mar\'{\i}a de Maeztu, grant MDM-2015-0509 and the Programa Severo
  Ochoa del Principado de Asturias; the Thalis and Aristeia programs
  cofinanced by EU-ESF and the Greek NSRF; the Rachadapisek Sompot
  Fund for Postdoctoral Fellowship, Chulalongkorn University and the
  Chulalongkorn Academic into Its 2nd Century Project Advancement
  Project (Thailand); the Kavli Foundation; the Nvidia Corporation;
  the SuperMicro Corporation; the Welch Foundation, contract C-1845;
  and the Weston Havens Foundation (USA).
\end{acknowledgments}

\section*{Data availability}
The authors will deposit their research data in accordance with the CMS data preservation, re-use and open access \href{https://cms-docdb.cern.ch/cgi-bin/PublicDocDB/RetrieveFile?docid=6032&filename=CMSDataPolicyV1.2.pdf&version=2}{policy}. Please also see  \href{http://cms-results.web.cern.ch/cms-results/public-results/publications/EXO-19-011}{the CMS Public Pages}.

\bibliography{auto_generated}
\cleardoublepage \appendix\section{The CMS Collaboration \label{app:collab}}\begin{sloppypar}\hyphenpenalty=5000\widowpenalty=500\clubpenalty=5000\vskip\cmsinstskip
\textbf{Yerevan Physics Institute, Yerevan, Armenia}\\*[0pt]
A.M.~Sirunyan$^{\textrm{\dag}}$, A.~Tumasyan
\vskip\cmsinstskip
\textbf{Institut f\"{u}r Hochenergiephysik, Wien, Austria}\\*[0pt]
W.~Adam, F.~Ambrogi, T.~Bergauer, M.~Dragicevic, J.~Er\"{o}, A.~Escalante~Del~Valle, M.~Flechl, R.~Fr\"{u}hwirth\cmsAuthorMark{1}, M.~Jeitler\cmsAuthorMark{1}, N.~Krammer, I.~Kr\"{a}tschmer, D.~Liko, T.~Madlener, I.~Mikulec, N.~Rad, J.~Schieck\cmsAuthorMark{1}, R.~Sch\"{o}fbeck, M.~Spanring, W.~Waltenberger, C.-E.~Wulz\cmsAuthorMark{1}, M.~Zarucki
% ...
\vskip\cmsinstskip
\dag: Deceased\\
1:  Also at Vienna University of Technology, Vienna, Austria\\
% ...
\end{sloppypar}
\end{document}